\documentclass[prd,floatfix,twocolumn]{revtex4}

\newcommand{\final}{1}

\usepackage{amsmath}  
\usepackage{amsfonts}  
\usepackage{graphicx}   

\newcommand{\myfig}[2]
{\centerline{\resizebox{!}{#1\textwidth}{\includegraphics{#2}}}}

\newcommand{\bex}{\begin{xalignat}{2}}
\newcommand{\be}{\begin{eqnarray}}
\newcommand{\ee}{\end{eqnarray}}
\newcommand{\ben}{\begin{eqnarray*}}
\newcommand{\een}{\end{eqnarray*}}
\newcommand{\eq}[1]{(\ref{#1})}
\newcommand{\degree}{$^\circ$}

{\newif\ifnotend
\notendtrue
\def\callist{ABCDEFGHIJKLMNOPQRSTUVWXYZ.}
\def\top#1#2.{#1}
\def\tail#1#2.{#2.}
\loop\expandafter\xdef\csname ff\expandafter\top\callist\endcsname%
{{\noexpand\sf\expandafter\top\callist}}
\edef\callist{\expandafter\tail\callist}
\if\callist.\notendfalse\fi\ifnotend\repeat}

{\newif\ifnotend
\notendtrue
\def\veclist{ABCDEFGHIJKLMNOPQRSTUVWXYZabcdefghijklmnopqrstuvwxyz.}
\def\top#1#2.{#1}
\def\tail#1#2.{#2.}
\loop\expandafter\xdef\csname v\expandafter\top\veclist\endcsname%
{{\noexpand\bf\expandafter\top\veclist}}
\edef\veclist{\expandafter\tail\veclist}
\if\veclist.\notendfalse\fi\ifnotend\repeat}

\newcommand{\boldsig}{\mbox{\boldmath $\sigma$}}

\newcommand{\boldrho}{\mbox{\boldmath $\rho$}}
\newcommand{\boldtheta}{\mbox{\boldmath $\theta$}}

\newcommand{\boldalpha}{\mbox{\boldmath $\alpha$}}

\newcommand{\LM}{{\cal M}}

\newcommand{\Psibar}{\overline{\Psi}}

\newcommand{\bis}{\mbox{\bfseries\itshape s}}
\newcommand{\bit}{\mbox{\bfseries\itshape t}}
\newcommand{\biu}{\mbox{\bfseries\itshape u}}
\newcommand{\biv}{\mbox{\bfseries\itshape v}}
\newcommand{\biw}{\mbox{\bfseries\itshape w}}

\newcommand{\ftS}{\mathbb{S}}

\newcommand{\bra}[1]{\left< {#1} \right|}
\newcommand{\ket}[1]{\left| {#1} \right>}
\newcommand{\braket}[2]{\left< \left. {#1} \right| {#2} \right>}
\newcommand{\Grad}{\mbox{\boldmath $\nabla$}}
\newcommand{\Div}{\mbox{\boldmath $\nabla$} \cdot}
\newcommand{\Curl}{\mbox{\boldmath $\nabla$} \wedge}

\newcommand{\half}{\mbox{\small $\frac{1}{2}$}}

\newcommand{\pd}[2]            
{ \left( \frac{ \partial {#1}}{\partial {#2}} \right)}
\newcommand{\lpd}[2]          
{ ( { \partial {#1}}/{\partial {#2}})}
\newcommand{\pdn}[2]            
{ \frac{ \partial {#1}}{\partial {#2}} }

\newcommand{\pp}{\partial}

\newcommand{\twovec}[2]{\left(\! \begin{array}{c} #1 \\ #2 \end{array}\!\right)}
\newcommand{\fourvec}[4]{\left(\! \begin{array}{c} #1\\#2\\#3\\#4 \end{array}\!\right)}
\newcommand{\fourmat}{ \left(\! \begin{array}{rrrr} }
\newcommand{\efourmat}{  \end{array} \! \right) }
\newcommand{\twomat}{ \left(\! \begin{array}{cc} }
\newcommand{\etwomat}{  \end{array} \! \right) }

\newcommand{\gap}[1]{ \rule{#1 cm}{0pt} }

\newcommand{\explainbox}[1]{\begin{table*}[t!] 
\rule{2 mm}{0 pt}\fbox{\hspace{1mm}
\begin{minipage}[t]{0.92\textwidth} \begin{flushleft}{#1}\end{flushleft} \end{minipage}
\hspace{1mm}}
\end{table*}}

\newcommand{\miniexplainbox}[1]{\begin{table}[t!] 
\rule{2 mm}{0 pt}\fbox{\hspace{1mm}
\begin{minipage}[t]{0.42\textwidth} \begin{flushleft}{#1}\end{flushleft} \end{minipage}
\hspace{1mm}}
\end{table}}

\newcommand{\eqbox}[3]
{ \fbox{
  \begin{minipage}[h]{0.43\textwidth}
    {\bf #1}
    \begin{eqnarray}
      #2       \label{#3}
    \end{eqnarray}
  \end{minipage}}
}

\begin{document}
\def\d{{\rm d}}
\newcounter{mycount}

\title{An introduction to spinors}

\author{Andrew M. Steane}
\affiliation{Department of Atomic and Laser Physics, Clarendon Laboratory, Parks Road, Oxford OX1 3PU, England.}

\date{\today}

\begin{abstract}
We introduce spinors, at a level appropriate for an undergraduate or first year graduate course
on relativity, astrophysics or particle physics. The treatment assumes very little mathematical
knowledge (mainly just vector analysis and some idea of what a group is). The SU(2)--SO(3)
homomorphism is presented in detail. Lorentz transformation, chirality, and the spinor Minkowski
metric are introduced. Applications to electromagnetism, parity violation, and to 
Dirac spinors are presented. A classical form of the Dirac equation is
obtained, and the (quantum) prediction that $g=2$ for Dirac particles is presented.
\end{abstract}





\maketitle

\section{Introducing spinors} \label{s.introspinor}

{\em Spinors} are mathematical entities somewhat like tensors, that allow
a more general treatment of the notion of invariance under rotation and Lorentz
boosts\footnote{This article has been prepared as a chapter of a textbook. A
preliminary, somewhat cut-down, version 
has been available on the author's web-site for a few years. It there began to
generate requests for further material, so it was decided to make it
more widely available.
Sources include \cite{52Payne,53Bade,98Feynman,12Steane,73Misner,96Ryder}
and some web-based material which I did not keep a note of. Starred sections can be omitted at first reading.}.
To every tensor of rank $k$ there corresponds a spinor of rank $2k$, and some kinds
of tensor can be associated with a spinor of the same rank. For example, a general 4-vector
would correspond to a Hermitian spinor of rank 2, which can be represented by a
$2 \times 2$ Hermitian matrix of complex numbers. A null 4-vector can also
be associated with a spinor of rank 1, which can be represented by a complex vector
with two components. We shall see why in the following.

Spinors can be used
without reference to relativity, but they arise naturally in discussions of
the Lorentz group. One could say that a spinor is the most basic sort of
mathematical object that can be Lorentz-transformed. The main facts about spinors
are given in the box on page \pageref{summarybox}.
The statements in the summary will be explained as we go along.


{\begin{table*}[ht!]    \label{summarybox}
\rule{2 mm}{0 pt}
\fbox{\hspace{1mm}
\begin{minipage}[t]{0.92\textwidth}

{
{\bf Spinor summary}.
A rank 1 spinor can be represented by a two-component complex vector, or
by a null 4-vector, angle and sign. The spatial part can be pictured as a flagpole with
a rigid flag attached.

The 4-vector is obtained from the 2-component complex vector by
\ben
\ffV^{\mu} &=  \bra{u} \sigma^{\mu} \ket{u} & \mbox{if $\biu$ is a contraspinor (``right-handed")}\\
\ffV_{\mu} &=  \bra{\tilde {u}} \sigma^{\mu} \ket{\tilde{u}} &
\mbox{if $\tilde{\biu}$ is a cospinor (``left handed")}.
\een

Any $2 \times 2$ matrix $\Lambda$ with unit determinant Lorentz-transforms a spinor.
Such matrices can be written
\[
\Lambda = \exp\left(i \boldsig \cdot \boldtheta/2 - \boldsig \cdot {\boldrho}/2 \right)
\]
where $\rho$ is rapidity. If $\Lambda$
is unitary the transformation is a rotation in space; if $\Lambda$ is Hermitian it is a boost.

If $\bis' = \Lambda(v) \bis$ is the Lorentz transform of a right-handed spinor, then under
the same change of reference frame a left-handed spinor transforms as
$\tilde{\bis}' = (\Lambda^{\dagger})^{-1} \tilde{\bis} = \Lambda(-v) \tilde{\bis}$.

The Weyl equations may be obtained by considering $(\ffW^{\alpha} \sigma_{\alpha}) \biw$.
This combination is zero in all frames. Applied to a spinor $\biw$ representing energy-momentum
it reads
\ben
(E/c - {\bf p} \cdot \boldsig) \biw &=& 0 \\
(E/c + {\bf p} \cdot \boldsig) {\tilde \biw} &=& 0.
\een
If we take $\sigma$ to represent spin angular momentum in these equations, then
the equations are not parity-invariant, and they imply that 
if both the energy-momentum and the spin of a particle can be represented simultaneously
by the same spinor, then the particle is massless and the sign of its helicity is fixed.

A Dirac spinor $\Psi  = (\phi_R, \phi_L)$ is composed of a pair of spinors, one
of each handedness. From the two associated null
4-vectors one can extract two orthogonal non-null 4-vectors
\ben
\ffV^{\mu} &=& \Psi^{\dagger}  \gamma^0 \gamma^{\mu} \Psi, \\
\ffW^{\mu} &=& \Psi^{\dagger}  \gamma^0 \gamma^{\mu} \gamma^5 \Psi,
\een
where $\gamma^{\mu}, \gamma^5$ are the Dirac matrices. With appropriate normalization factors
these 4-vectors can represent the 4-velocity and 4-spin of a particle such as the electron.

Starting from a frame in which $\ffV^i = 0$ (i.e. the rest frame), the result of a Lorentz boost
to a general frame can be written
\[
\twomat -m & E + \boldsig \cdot {\bf p} \\
E - \boldsig \cdot {\bf p} & -m  \etwomat \twovec{\phi_R({\bf p})}{\chi_L({\bf p})} = 0.
\]
This is the Dirac equation. Under parity inversion the parts of a Dirac spinor swap over
and $\boldsig \rightarrow -\boldsig$; the Dirac
equation is therefore parity-invariant.

}
\end{minipage}
\hspace{1mm}}
\end{table*}}

It appears that Klein originally designed the spinor to
simplify the treatment of the classical spinning top in 1897. The more thorough
understanding of spinors as mathematical objects is credited to
\'{E}lie Cartan in 1913. They are closely related to Hamilton's quaternions (about 1845).

Spinors began to find a more extensive role in physics when it was
discovered that electrons and other particles
have an intrinsic form of angular momentum now called `spin', and the behaviour
of this type of angular momentum is correctly captured by the mathematics
discovered by Cartan. Pauli formalized this connection in a non-relativistic
(i.e. low velocity) context, modelling the electron spin using a two-component
complex vector, and introducing the {\em Pauli spin matrices}.
Then in seeking a quantum mechanical description of
the electron that was consistent with the requirements of Lorentz covariance,
Paul Dirac had the brilliant insight that an equation of the right form could
be found if the electron is described by combining the mathematics of spinors
with the existing quantum mechanics of wavefunctions. He introduced
a 4-component complex vector, now called a {\em Dirac spinor}, and by
physically interpreting the wave equation thus obtained, he
predicted the existence of antimatter.


Here we will discuss spinors in general, concentrating on the simplest case, namely 2-component
spinors. These suffice to describe rotations
in 3 dimensions, and Lorentz transformations in $3+1$ dimensions. We will
briefly introduce the spinors of higher rank, which transform like outer products of
first rank spinors. We will then introduce Dirac's idea, which can be
understood as a pair of coupled equations for a pair of first rank spinors.

Undergraduate students often first meet spinors in the context of non-relativistic
quantum mechanics and the treatment of the spin angular momentum. This can give
the impression that spinors are essentially about spin, an impression that is
fortified by the name `spinor'. However, you should try to avoid that assumption
in the first instance. Think of the word `spinor' as a generalisation of `vector'
or `tensor'. We shall meet a spinor that describes an electric 4-current, for example,
and a spinor version of the Faraday tensor, and
thus write Maxwell's equations in spinor notation.

\begin{figure}
\myfig{0.30}{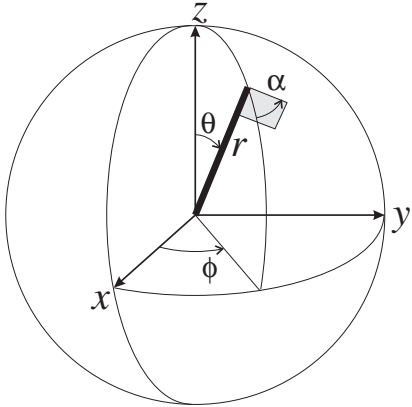}
\caption{A spinor. The spinor has a direction in space (`flagpole'), an orientation
about this axis (`flag'), and an overall sign (not shown). A suitable set of parameters
to describe the spinor state, up to a sign, is $(r,\theta,\phi,\alpha)$, as shown. The
first three fix the length and direction of the flagpole by using standard spherical
coordinates, the last gives the orientation of the flag.}
\label{f.spinor}
\end{figure}

Just as we can usefully think of a vector as an arrow in space,
and a 4-vector as an arrow in spacetime, it is useful to have a geometrical
picture of a rank 1 spinor (or just `spinor' for short);  see
figure \ref{f.spinor}.
It can be pictured as a vector with two further features:
a `flag' that picks out a plane in space containing the vector, and an overall
sign. The crucial property is that under the action of a rotation,
the direction of the spinor changes just as a vector would, and the flag is carried along
in the same way is if it were rigidly attached to the `flag pole'. A rotation
about the axis picked out by the flagpole would have no effect on a vector pointing
in that direction, but it does affect the spinor because it rotates the flag.

The overall sign of the spinor is more subtle. We shall find that when a spinor is rotated
through 360$^{\circ}$, it is returned to its original direction, as one would expect, but also
it picks up an overall sign change. You can think of this as a phase factor ($e^{i \pi} = -1$).
This sign has no consequence when spinors are examined one at a time, but it can be relevant
when one spinor is compared with another. When we introduce the mathematical description using
a pair of complex numbers (a 2-component complex vector) this and all other properties will
automatically be taken into account.


To specify a spinor state one must furnish 4 real parameters and a sign: an illustrative
set $r,\theta,\phi,\alpha$ is given in figure \ref{f.spinor}. One can see that just such
a set would be naturally suggested if one wanted to analyse the motion of a spinning top.
We shall assume the overall sign is positive unless explicitly stated otherwise.
The application to a classical spinning top is such that the spinor could represent
the instantaneous positional state of the top. However, we shall not be interested
in that application. In this article we will show how a spinor can be used to represent
the energy-momentum and the spin of a massless particle, and a pair of spinors
can be used to represent the energy-momentum and Pauli-Lubanksi spin 4-vector
of a massive particle. Some very interesting properties of spin angular
momentum, that otherwise might
seem mysterious, emerge naturally when we use spinors.

A spinor, like a vector, can be rotated.
Under the action of a rotation, the spinor magnitude is fixed while the angles $\theta,\phi,\alpha$
change. In the flag picture, the flagpole and flag evolve together as a rigid body; this
suffices to determine
how $\alpha$ changes along with $\theta$ and $\phi$. In order to write
the equations determining the effect of a rotation, it is convenient
to gather together the four parameters into two complex numbers defined by
\be
a &\equiv& \sqrt{r} \cos(\theta/2) e^{i(-\alpha-\phi)/2},  \nonumber \\
b &\equiv& \sqrt{r} \sin(\theta/2) e^{i(-\alpha+\phi)/2}.  \label{def_ab}
\ee
(The reason for the square root and the factors of 2 will emerge in the discussion).
Then the effect of a rotation of the spinor through $\theta_r$ about the $y$ axis, for example, is
\be
\twovec{a'}{b'} = \twomat \cos(\theta_r/2) & -\sin(\theta_r/2)
                           \\ \sin(\theta_r/2) & \;\;\;\cos(\theta_r/2) \etwomat
                           \twovec{a}{b}.         \label{eg_yrotate}
\ee
We shall prove this when we investigate more general rotations below.

From now on we shall
refer to the two-component complex vector
\be
\bis = s e^{-i \alpha/2} \twovec{ \cos(\theta/2) e^{-i\phi/2} }
                           {\sin(\theta/2) e^{i\phi/2}}  \label{defspinor}
\ee
as a `spinor'.
A spinor of size $s$ has a flagpole of length
\be
r = |a|^2 + |b|^2 = s^2.      \label{r_from_s}
\ee
The components $(r_x, r_y, r_z)$ of the flagpole vector are given by
\be
r_x = a b^* + b a^*, \;\; r_y = i (a b^* - b a^*), \;\; r_z = |a|^2 - |b|^2,         \label{nxnynz}
\ee
which may be obtained by inverting (\ref{def_ab}). You can now see why the square root
was required in (\ref{def_ab}). 

The complex number representation will prove to be central to understanding spinors.
It gives a second picture of a spinor, as a vector in a 2-dimensional complex vector
space. One learns to `hold' this picture alongside the first one.
Most people find themselves
thinking pictorially in terms of a flag in a 3-dimensional real space as illustrated in
figure \ref{f.spinor}, but every now and then it is helpful to remind oneself that a
pair of opposite flagpole states such
as `straight up along $z$' and `straight down along $z$' are
{\em orthogonal} to one another in the complex vector space (you can see this
from eq. (\ref{defspinor}), which gives $(s,0)$ and $(0,s)$ for these cases,
up to phase factors).

\begin{figure}
\myfig{0.35}{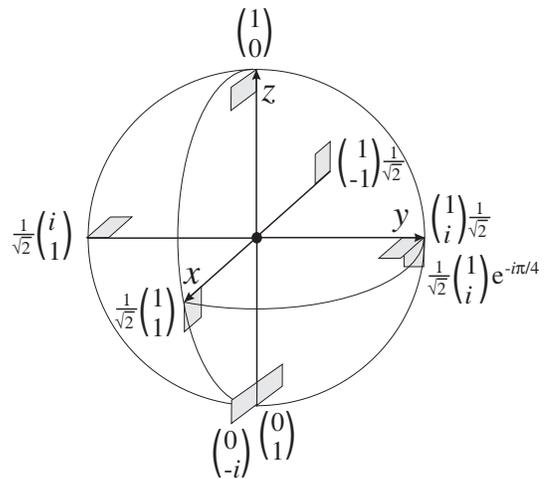}
\caption{Some example spinors. In two cases a pair of
spinors pointing in the same direction but with flags in different directions
are shown, to illustrate the role of the flag angle $\alpha$. Any given direction
and flag angle can also be represented by a spinor of opposite sign to the one
shown here.}
\label{f.spinoreg}
\end{figure}

Figure \ref{f.spinoreg} gives some example spinor states with their complex
number representation. Note that the two basis vectors $(1,0)$ and $(0,1)$ are
associated with flagpole directions up and down along $z$, respectively, as we
just mentioned. Considered
as complex vectors, these are orthogonal to one another, but they represent
directions in 3-space that are opposite to one another. More generally,
a rotation through an angle
$\theta_r$ in the complex `spin space' corresponds to a rotation through an angle
$2\theta_r$ in the 3-dimensional real space.
This is called `angle doubling'; you can see it in eq (\ref{defspinor}) and we shall
explore it further in section \ref{s.su2}.

The matrix (\ref{eg_yrotate}) for rotations about the $y$ axis is real, so spinor
states obtained by rotation of $(1,0)$ about the $y$ axis are real. These all have the
flag and flagpole in the $xz$ plane, with the flag pointing in the right handed
direction relative to the $y$ axis (i.e. the clockwise direction when the $y$ axis
is directed into the page). A rotation about the $z$ axis is represented by a diagonal
matrix, so that it leaves spinor states $(1,0)$ and $(0,1)$ unchanged in direction. To
find the diagonal matrix, consider the spinor $(1,1)$ which is directed
along the positive $x$ axis. A rotation about $z$ should increase $\phi$ by the
rotation angle $\theta_r$. This means the matrix for
a rotation of the spinor about the $z$ axis through angle $\theta_r$ is
\be
 \twomat \exp(-i\theta_r/2) & 0 \\
          0 & \exp(i\theta_r/2) \etwomat.          \label{eg_zrotate}
\ee
When applied to the spinor $(1,0)$, the result is $(e^{-i \theta_r/2},0)$.
This shows that the result is to increase $\alpha+\phi$ from zero to
$\theta_r$. Therefore the flag is rotated. In order to be consistent with
rotations of spinor directions close to the $z$ axis, it makes sense to
interpret this as a change in $\phi$ while leaving $\alpha$ unchanged.

So far our spinor picture was purely a spatial one. We are used to putting
3-vectors into spacetime by finding a fourth quantity and forming a 4-vector.
For the spinor, however, a different approach is used, because it will turn out
that the spinor is already a spacetime object that can be Lorentz-transformed.
To `place' the spinor in spacetime we just need to identify the 3-dimensional
region or `hypersurface' on which it lives. We will show in section \ref{s.4vecspin}
that the 4-vector associated with the flagpole is a null 4-vector. Therefore,
the spinor should be regarded as `pointing along' or `existing on' the light cone.
The word `cone' suggests a two-dimensional surface, but of course it is 3-dimensional really
and therefore can contain a spinor.
The event whose light cone is meant will be clear in practice. For example,
if a particle has mass or charge then we say the mass or charge is located
at each event where the particle is present. In a similar way, if a
rank 1 spinor is used to describe a property of a particle, then the spinor can
be thought of as `located at' each event where the particle is present, and
lying on the future light cone of the event.
(Some spinors of higher rank can also be associated with
4-vectors, not necessarily null ones.)
The formula for a null 4-vector,
$(\ffX^0)^2 = (\ffX^1)^2 + (\ffX^2)^2 + (\ffX^3)^2$, leaves open a
choice of sign between the time and spatial parts, like the distinction
between a contravariant and covariant 4-vector. We shall show in section
\ref{s.chirality} that this
choice leads to two types of spinor, called `left handed' and `right handed'.


\section{The rotation group and SU(2)}    \label{s.su2}

We introduced spinors above by giving a geometrical picture, with flagpole and flag
and angles in space. We then gave another definition, a 2-component complex vector.
We have an equation relating the definitions, (\ref{def_ab}). All this makes
it self-evident that there must exist a set of transformations of the complex vector
that correspond to {\em rotations} of the flag and flagpole. It is also easy to
guess what transformations these are: they have to preserve the length $r$ of the flagpole, so
they have to preserve the size $|a|^2 + |b|^2$ of the complex vector. This implies they
are {\em unitary} transformations. If you are happy to accept this, and if you are happy to
accept eq. (\ref{summarizeUandR}) or prove it by others means (such as trigonometry), then you
can skip this section and proceed straight to section \ref{s.eigen}. However,
the connection between rotations and unitary $2 \times 2$ matrices gives an important example of a very
powerful idea in mathematical physics, so in this section we shall take some trouble to explore it.

The basic idea is to show that two groups, which are defined in different ways in
the first instance, are in fact the same (they are in one-to-one correspondance with one another, called
isomorphic) or else very similar (e.g. each element of one group corresponds
to a distinct set of elements of the other, called homomorphic). These are mathematical
{\em groups} defined as in group theory, having associativity, closure, an identity element and
inverses. The groups we are concerned with have a continuous range of members, so are called Lie groups.
We shall establish one of the most important
mappings in Lie group theory (that is, important to physics---mathematicians would
regard it as a rather simple example). This is the `homomorphism'
\be
 \mbox{SU(2)} \stackrel{2:1}{\longrightarrow} \mbox{S0(3)}
\ee
`Homomorphism' means the mapping is not one-to-one; here there are two elements
of SU(2) corresponding to each element of SO(3). SU(2) is the {\em special unitary
group of degree 2}. This is the group of two by two unitary\footnote{A {\em unitary}
matrix is one whose Hermitian conjugate is its inverse, i.e. $U U^{\dagger} = I$.}
matrices with determinant~1. SO(3) is the
special orthogonal group of degree 3, isomorphic to the {\em rotation group}.
The former is the group of
three by three orthogonal\footnote{An {\em orthogonal} matrix is
one whose transpose is its inverse, i.e. $R R^T = I$.}
real matrices with determinant 1.
The latter is the group of rotations about the origin
in Euclidian space in 3 dimensions.

These Lie groups SU(2) and SO(3) have the same `dimension', where the dimension
counts the number of real parameters needed to specify a member of the group.
This `dimension' is the number of dimensions of the abstract `space' (or {\em manifold})
of group members (do not confuse this with dimensions in physical space and time). The rotation group
is three dimensional because three parameters are needed to specify a rotation
(two to pick an axis, one to give the rotation angle); the matrix group SO(3)
is three dimensional because a general $3 \times 3$ matrix has nine parameters,
but the orthogonality and unit determinant conditions together set six constraints;
the matrix group SU(2) is three dimensional because a general $2 \times 2$
unitary matrix can be described
by 4 real parameters (see below) and the determinant condition gives one constraint.

The strict definition of an isomorphism between groups is as follows. If $\{M_i\}$
are elements of one group and $\{N_i\}$ are elements of the other, the groups
are isomorphic if there exists a mapping $M_i \leftrightarrow N_i$ such that
if $M_i M_j = M_k$ then $N_i N_j = N_k$. For a homomorphism the same condition
applies but now the mapping need not be one-to-one.

The SU(2), SO(3) mapping may be established as follows.
First introduce the {\em Pauli spin matrices}
\[
\sigma_x = \twomat 0 & 1 \\ 1 & 0 \etwomat, \;
\sigma_y = \twomat 0 & -i \\ i & 0 \etwomat, \;
\sigma_z = \twomat 1 & 0 \\ 0 & -1 \etwomat.
\]
Note that these are all Hermitian and unitary. It follows that they square to one:
\be
\sigma_x^2 = \sigma_y^2 = \sigma_z^2 = I.
\ee
They also have zero trace. It is very useful to know their {\em commutation relations}:
\be
\left[\sigma_x,\, \sigma_y\right] \equiv \sigma_x \sigma_y - \sigma_y \sigma_x = 2 i \sigma_z
    \label{sigmacommute}
\ee
and similarly for cyclic permutation of $x,y,z$. You can also notice that
\[
\sigma_x \sigma_y = i \sigma_z, \;\; \sigma_y \sigma_x = - i \sigma_z
\]
and therefore any pair anti-commutes:
\be
\sigma_x \sigma_y = -\sigma_y \sigma_x   \label{xyanti}
\ee
or in terms of the `anticommutator'
\be
\{ \sigma_x, \sigma_y \} \equiv \sigma_x \sigma_y + \sigma_y \sigma_x = 0.
\ee

Now, for any given spinor $\bis$, the components of the flagpole
vector, as given by eqn (\ref{nxnynz}), can be written
\be
r_x =  \bis^{\dagger} \sigma_x \bis, \; 
r_y =  \bis^{\dagger} \sigma_y \bis, \; 
r_z =  \bis^{\dagger} \sigma_z \bis
\ee
which can be written more succinctly,
\be
\vr =  \bis^{\dagger} \boldsig \bis  = \bra{s} \boldsig \ket{s}
\ee
where the second version is in Dirac notation\footnote{We will
occasionally exhibit Dirac notation alongside the
vector and matrix notation, for the benefit of readers familiar with it.
If you are not such a reader than you can safely ignore this. It is easily
recognisable by the presence of $\left< \,|\, \right>$ angle bracket symbols.}.

Consider (exercise \ref{ex.expsig})
\be
e^{i(\theta/2) \sigma_j} &=&  \cos(\theta/2) I + i \sin(\theta/2) \sigma_j
\label{expsigma}
\ee
hence
\be
e^{i(\theta/2) \sigma_x} &=& 
\twomat \cos(\theta/2) & i \sin(\theta/2)
                           \\ i \sin(\theta/2) & \cos(\theta/2) \etwomat,  \label{rotsigx} \\
e^{i(\theta/2) \sigma_y} &=& 
\twomat \cos(\theta/2) & \sin(\theta/2)
                           \\ -\sin(\theta/2) & \cos(\theta/2) \etwomat,  \label{rotsigy} \\
e^{i(\theta/2) \sigma_z} &=& 
\twomat e^{i\theta/2} & 0 \\ 0 & e^{-i \theta/2}\etwomat.    \label{rotsigz}
\ee
We shall call these the `spin rotation matrices.'
We will now show that when the spinor is acted on by the matrix
$\exp(i \theta \sigma_x/2)$, the flagpole is rotated through the angle $\theta$
about the $x$-axis. This can be shown directly from
eq. (\ref{defspinor}) by trigonometry, but it will be more
instructive to prove it using matrix methods, as follows. Let
\[
\bis' = e^{i(\theta/2) \sigma_x} \bis
\]
then
\be
\vr' = \bra{s'}\boldsig\ket{s'} 
= \bra{s}  e^{-i(\theta/2) \sigma_x}\,\boldsig\, e^{i(\theta/2) \sigma_x}\ket{s}
\label{rprime}
\ee
where we used that $\sigma_x$ is Hermitian. Consider first the $x$-component of this expression:
\be
x' &=& \bra{s}  e^{-i(\theta/2) \sigma_x}\sigma_x e^{i(\theta/2) \sigma_x}\ket{s}  \nonumber\\
&=& \bra{s}  \sigma_x e^{-i(\theta/2)\sigma_x} e^{i(\theta/2) \sigma_x}\ket{s} \nonumber\\
&=& \bra{s}  \sigma_x \ket{s} = x,
\ee
where we used that $\sigma_x$ commutes with $I$ and itself, and you should confirm
that 
\be
e^{-i(\theta/2)\sigma_x} e^{i(\theta/2) \sigma_x} = I
\ee
(more generally, if a matrix $H$ is Hermitian then $\exp(iH)$ is unitary). Now consider the $y$-component
of (\ref{rprime}):
\be
y' &=&\bra{s}  e^{-i(\theta/2) \sigma_x}\sigma_y e^{i(\theta/2) \sigma_x}\ket{s}  \label{yprime}
\ee
To reduce clutter in the following, introduce $\alpha = \theta/2$. Then,
using (\ref{expsigma}), the operator in the middle of (\ref{yprime}) is
\be
(\cos\alpha \!&-&\!i\sin\alpha \sigma_x) \sigma_y (\cos\alpha +i\sin\alpha \sigma_x)  \nonumber \\
&=& \sigma_y (\cos\alpha  + i\sin\alpha \sigma_x) (\cos\alpha +i\sin\alpha \sigma_x) \nonumber \\
&=& \sigma_y (\cos^2\alpha  - \sin^2\alpha + 2 i \sin\alpha \cos\alpha  \sigma_x ) \nonumber \\
&=& \sigma_y ( \cos \theta + i \sin\theta  \sigma_x )  \label{opy}
\ee
where in the first step we brought $\sigma_y$ to the front by using 
that $\sigma_x$ and $\sigma_y$ anti-commute (eqn (\ref{xyanti})), and in the second
step we used that $\sigma_x^2 = I$. Upon subsituting the result (\ref{opy}) into (\ref{yprime})
we have
\be
y' = \cos \theta \bra{s}\sigma_y\ket{s} + i\sin\theta\bra{s}\sigma_y\sigma_x\ket{s}
\ee
but $\sigma_y \sigma_x = -i \sigma_z$, so this is
\be
y' = \cos(\theta) y + \sin(\theta) z.
\ee
The analysis for $z'$ goes the same, except that we have $\sigma_z \sigma_x = +i\sigma_y$
in the final step, so
\be
z' = \cos(\theta) z - \sin(\theta) y.
\ee
The overall result is $\vr' = R_x \vr$, where $R_x$ is the matrix representing a rotation through
$\theta$ about the $x$ axis. Owing to the fact that the commutation relations (\ref{sigmacommute})
are obeyed by cyclic permutations of $x,y,z$, the corresponding results for $\sigma_y$ and $\sigma_z$
immediately follow. Therefore, we have shown that multiplying a spinor by each of the
spin rotation matrices (\ref{rotsigx})-(\ref{rotsigz}) results in a rotation of the flagpole by
the corresponding matrix for a rotation in three dimensions:
\be
R_x &=& \left( \begin{array}{rrr} 1 & 0 & 0 \\
0 & \cos \theta & \sin \theta \\
0 & -\sin \theta & \cos \theta
\end{array} \right)        \\
R_y &=& \left( \begin{array}{rrr} \cos \theta & 0 & -\sin \theta \\
0 & 1 & 0 \\
\sin \theta & 0 & \cos \theta
\end{array} \right)     \label{Ry_rot} \\
R_z &=& \left( \begin{array}{rrr}
\cos \theta & +\sin \theta & 0 \\
-\sin \theta & \cos \theta & 0 \\
0 & 0 & 1
\end{array} \right).    \label{Rz_rot}
\ee
These are rotations about the $x$, $y$ and $z$ axes respectively, but note the {\em angle
doubling}: the rotation angle $\theta$ is twice the angle $\theta/2$ which appears
in the $2 \times 2$ `spin rotation' matrices. The sense of rotation
is such that $R$ represents a change of reference frame, that is to say, a rotation of
the coordinate axes in a right-handed sense\footnote{This is why (\ref{rotsigy})
and (\ref{rotsigz}) have a rotation angle of opposite sign to (\ref{eg_yrotate}) and
(\ref{eg_zrotate}).}.

We have now almost established the homomorphism between the groups
SU(2) and SO(3), because we have explicitly stated which member of SU(2)
corresponds to which member of SO(3). It only remains to note that {\em any}
member of SU(2) can be written (exercise \ref{ex.unitary})
\be
U = e^{i \boldsig \cdot \boldtheta}
\ee
and {\em any} member of SO(3) can be written
\[
R = e^{i \vJ \cdot \boldtheta}
\]
where $\vJ$ are the generators of rotations in three dimensions:
\be
J_x &=& \left( \begin{array}{rrr} 0 & 0 & 0 \\ 0 & 0 & -i \\ 0 & i & 0 \end{array}\right), \nonumber \\
J_y &=& \left( \begin{array}{rrr} 0 & 0 & i \\ 0 & 0 & 0 \\ -i & 0 & 0 \end{array}\right), \nonumber \\
J_z &=& \left( \begin{array}{rrr} 0 & -i & 0 \\ i & 0 & 0 \\ 0 & 0 & 0 \end{array}\right).
                   \label{Jgenerator}
\ee
Note also that to obtain a given rotation $R$, we can use either $U$ or $-U$. We have now
fully established the mapping between the groups:

\vspace{6pt}
\fbox
{
  \begin{minipage}[h]{0.41\textwidth}
\begin{tabular}{lcl}
spinor $\bis$  & $\leftrightarrow$ & vector $\vr =\bra{s}\boldsig\ket{s}$  \\
\hline
\begin{tabular}{l}
Members $U$  \\
and $-U$ of SU(2) \end{tabular}
&$\leftrightarrow$& member $R$ of SO(3)\\
\hline
$U = e^{i \boldsig \cdot \boldtheta/2}$ \rule{0pt}{3.5ex} & & $R = e^{i {\bf J} \cdot \boldtheta}$
\end{tabular}
  \end{minipage}
}
\be  \label{summarizeUandR} \ee

Let us note also the effect of an inversion of the coordinate system through the origin (called
parity inversion). Vectors such as displacement change sign under such an inversion and are
called {\em polar vectors}. Vectors such as angular momentum do not change sign under
such an inversion, and are called {\em axial vectors} or {\em pseudovectors}.
Suppose polar vectors $\va$ and $\vb$ are related
by $\vb = R \va$. Under parity
inversion these vectors transform as $\va' = -\va$ and $\vb'=-\vb$, so one finds
$\vb' = R \va'$, hence the rotation matrix is unaffected by parity inversion: $R' = R$. It follows
that, in the expression $R = \exp(i \vJ \cdot \boldsig)$, we must take $\vJ$ and $\boldsig$
as either both polar or both axial. The choice, whether we consider $\boldsig$ to be polar
or axial, depends on the context in which it is being used.

With the benefit of hindsight, or else with a good knowledge of group theory,
one could `spot' the SU(2)--SO(3) homomorphism, including the angle doubling, simply by noticing
that the commutation relations (\ref{sigmacommute}) are the same as those
for the rotation matrices $J_i$, apart from the factor 2.  This is because, if you
look back through the argument, you can see that it would apply to any set
of quantities obeying those relations. More generally, therefore, we say that
the Pauli matrices are defined to be a set of entities that obey the commutation relations,
and their standard expressions using $2\times 2$ matrices are 
one {\em representation} of them.

The angle doubling leads to the curious feature that when $\theta = 2\pi$
(a single full rotation) the spin rotation matrices all give $-I$. It is {\em not} that
the flagpole reverses direction---it does not, and neither does the flag---but rather, the
spinor picks up an overall sign that has no ready representation in the flagpole picture.

\begin{figure}
\myfig{0.2}{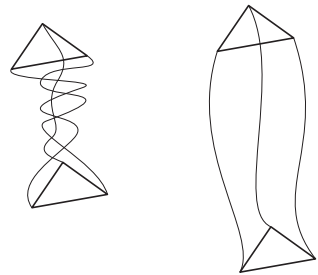}
\caption{`Tangloids' is a game invented by Piet Hein to explore the effect of rotations
of connected objects. Two small wooden poles or triangular blocks are joined by three parallel strings. Each
player holds one of the blocks. The first player holds one block still, while the other
player {\em rotates the other wooden block for two full revolutions about any fixed axis}.
After this, the strings appear to be tangled. The first player now has to untangle them
{\em without} rotating either piece of wood. He must use a parallel transport, that is,
a translation of his block (in 3 dimensions) without rotating it or the other block.
The fact that it can be done (for a 720$^{\circ}$
initial rotation, but not for a 360$^{\circ}$ initial rotation) illustrates a subtle
property of rotations. After swapping roles, the winner is the one who untangled
the fastest.}
\label{f.tangloids}
\end{figure}

It is worth considering for a moment. We usually consider
that a 360$^{\circ}$ leaves everything unchanged. This is true for a global
rotation of the whole universe, or for a rotation of an isolated object not
interacting with anything else. However, when one object is rotated while interacting
with another that is not rotated, more possibilities arise. The fact that a spinor rotation
through 360$^{\circ}$ does not give the identity operation captures a valid
property of rotations that is
simply not modelled by the behaviour of vectors. Place a fragile object such as a china plate on
the palm of your hand, and then rotate your palm through 360$^{\circ}$ (say,
anticlockwise if you use your right hand) while keeping
your palm horizontal, with the plate balanced on it. It can
be done but you will now be standing somewhat awkwardly with a twist in your arm.
But now {\em continue} to rotate your palm {\em in the same direction} (still
anticlockwise). It can be done: most of us find ourselves bringing our hand up
over our shoulder, but note: the palm and plate remain horizontal and continue
to rotate. After thus completing two full revolutions, 720$^{\circ}$, you should find yourself
standing comfortably, with no twist in your arm! This simple experiment illustrates
the fact that there is more to rotations than is captured by the simple notion
of a direction in space. Mathematically, it is noticed in a subtle property of
the Lie group SO(3): the associated smooth space is not `simply connected'
(in a topological sense). The group SU(2) exhibits it more clearly:
the result of one full rotation is a sign change; a second full rotation is
required to get a total effect equal to the identity matrix. Figure \ref{f.tangloids}
gives a further comment on this property.

\subsection{Rotations of rank 2 spinors}  \label{rank2}

The mapping between SU(2) and SO(3) can also be established by examining a class
of second rank spinors. This serves to introduce some further useful ideas.

For any real vector ${\bf r} = (x,y,z)$ one can construct the traceless
Hermitian matrix
\be
  X = {\bf r} \cdot \boldsig = x \sigma_x + y \sigma_y + z \sigma_z =
      \twomat z & x - i y \\ x + iy & -z  \etwomat.              \label{Xrdotsig}
\ee
It has determinant
\[
|X| = -(x^2 + y^2 + z^2).
\]
Now consider the matrix product
\be
U X U^{\dagger} = X'          \label{UXU}
\ee
where $U$ is unitary and of unit determinant.
For {\em any} unitary $U$, if $X$ is Hermitian then the result $X'$ is (i) Hermitian
and (ii) has the same trace. Proof (i):
$(X')^{\dagger} = (U X U^{\dagger})^{\dagger} = (U^{\dagger})^{\dagger} X^{\dagger}
U^{\dagger} = U X U^{\dagger} = X'$; (ii): the trace is the sum of the eigenvalues
and the eigenvalues are preserved in unitary transformations.
Since $X'$ is Hermitian and traceless, it can in turn be interpreted as
a 3-component real vector ${\bf r}'$ (you are invited to prove this after reading on),
and furthermore, if $U$ has determinant 1 then $X'$ has the same determinant as $X$
so ${\bf r}'$ has the same length as $\vr$.
It follows that the transformation of $\vr$ is either a rotation or a reflection.
We shall prove that it is a rotation. To do this, it suffices to pick one of
the spin rotation matrices; for convenience choose the $z$-rotation:
\[
e^{i(\theta/2)\sigma_z} X e^{-i(\theta/2)\sigma_z} =
\twomat
z  & e^{i\theta}(x-iy) \\ e^{-i\theta} (x+iy) & -z \etwomat .
\]
The vector associated with this matrix is $(x \cos\theta + y\sin\theta,\;
-x \sin\theta + y\cos\theta, z)$, which is $R_z \vr$.

The relationship between the groups follows as before.

\subsection{Spinors as eigenvectors}  \label{s.eigen}

In this section we present an idea which is much used in quantum physics, but also
has wider application because it is part of the basic mathematics of spinors.
Every vector can be considered to be the eigenvector, with eigenvalue 1,
of an orthogonal matrix, and a similar property applies to spinors.
We will show that every spinor is the eigenvector,
with eigenvalue 1, of a $2\times 2$ traceless Hermitian matrix. But, we saw in the previous section that
such matrices can be related to vectors, so we have another interesting connection.
It will turn out that the direction associated with the matrix will agree with the flagpole
direction of the spinor!

In fact, it is this result that motivated the assignment that we started with, eqn (\ref{def_ab}).
The proofs connecting SU(2) matrices to SO(3) matrices do not themselves require any
particular choice of the assignment of 3-vector direction to a complex 2-vector (spinor), 
only that it be assigned in a way that makes sense when rotations are applied. After all,
we connected the groups in section \ref{rank2} without mentioning rank 1 spinors at all.
The choice (\ref{def_ab}) either leads to, or, depending on your point of view, follows
from, the considerations we are about to presemt.

{\em Proof.} First we show that for any 2-component complex
vector $\bis$ we can construct a matrix
$S$ such that $\bis$ is an eigenvector of $S$ with
eigenvalue 1. 

We would like $S$ to be Hermitian. To achieve this,
we make sure the eigenvectors are orthogonal and the eigenvectors real.
The orthogonality we have in mind here is with respect to the standard definition of
inner product in a complex vector space, namely
\[
\biu^{\dagger} \biv = u_1^* v_1 + u_2^* v_2 = \braket{u}{v}
\]
where the last version on the right is in Dirac notation\footnote{We will
occasionally exhibit Dirac notation alongside the
vector and matrix notation, for the benefit of readers familiar with it.
If you are not such a reader than you can safely ignore this. It is easily
recognisable by the presence of $\left< \,|\, \right>$ angle bracket symbols.}.
Beware, however, that we shall be introducing another type of inner product
for spinors in section \ref{s.chirality}.

Let $\bis = \twovec{a}{b}$. The spinor orthogonal\footnote{The two
eigenvectors, considered as vectors in a complex
vector space, are orthogonal to one another (because $S_{\bf n}$ is Hermitian), but their
associated flagpole directions are opposite. This is an example of the angle doubling
we already noted in the relationship between SU(2) and SO(3).}
to $\bis$ and with the same length is
$\twovec{-b^*}{a^*}$ (or a phase factor times this).
Let the eigenvalues be $\pm 1$, then we have
\[
S V = V \sigma_z
\]
where
\[
V  = \frac{1}{s}\twomat a & -b^* \\ b & a^* \etwomat
\]
is the matrix of normalized eigenvectors, with $s = \sqrt{|a|^2 + |b|^2}$.
$V$ is unitary when the eigenvectors are normalized, as
here. The solution is
\be
S = V \sigma_z V^{\dagger} = \frac{1}{s^2} \twomat |a|^2 - |b|^2 & 2 a b^* \\
2 b a^* & |b|^2 - |a|^2 \etwomat.  \label{Sfroms}
\ee
Comparing this with (\ref{Xrdotsig}), we see that the direction associated with $S$
is as given by (\ref{nxnynz}). Therefore {\em the direction associated with
the matrix $S$ according to (\ref{Xrdotsig}) is the same as the flagpole direction of
the spinor $\bis$ which is an
eigenvector of $S$ with eigenvalue 1. QED.}

Eq. (\ref{Sfroms}) can be written
\[
S = {\bf n} \cdot \boldsig = n_x \sigma_x + n_y \sigma_y + n_z \sigma_z
\]
where $n_x$, $n_y$, $n_z$ are given by equations (\ref{nxnynz}) divided by $s^2$.
We find that $\bf n$ is a unit vector (this comes from the choice that the
eigenvalue is 1). The result can also be written
$n_x = \bis^{\dagger} \sigma_x \bis/s^2$ and similarly.
More succinctly, it is
\be
{\bf n} = \frac{\bis^{\dagger} \boldsig \bis }{s^2} = \frac{\bra{s} \boldsig \ket{s}}{s^2},
      \label{meansigma}
\ee

Another useful way of stating the overall conclusion is
\begin{quote}
{For any unit vector $\bf n$, the Hermitian traceless matrix
\[
S ={\bf n} \cdot {\boldsig}
\]
has an eigenvector of eigenvalue 1 whose flagpole is along $\bf n$.}
\end{quote}

Since a rotation of the coordinate system would bring $S$ onto one
of the Pauli matrices, $S$ is called a `spin matrix' for spin along
the direction $\bf n$.

Suppose now that we have another spinor related to the first one
by a rotation: $\bis' = U \bis$. We
ask the question, of which matrix is $\bis'$ an eigenvector
with eigenvalue 1? We propose and verify the solution $U S U^{\dagger}$:
\[
(U S U^{\dagger}) \bis' = U S U^{\dagger} U \bis
= U S \bis = U \bis = \bis'.
\]
Therefore the answer is
\[
S' = U S U^{\dagger}.
\]
This is precisely the transformation that represents a
rotation of the vector ${\bf n}$ (compare with (\ref{UXU})),
so we have proved that the flagpole of $\bis'$ is in the direction $R {\bf n}$,
where $R$ is the rotation in 3-space associated with $U$
in the mapping between SU(2) and SO(3).
Therefore $U$ gives a rotation of the direction of the spinor.

We have here presented spinors as classical (in the sense of not quantum-mechanical)
objects. If you suspect that the occasional mention of Dirac notation means that we
are doing quantum mechanics, then please reject that impression. 
in this article a spinor is a classical object. It is a generalization of a classical vector.

\section{Lorentz transformation of spinors}  \label{s.lorentz_spinor}

We are now ready to generalize from space to spacetime, and make contact
with Special Relativity. It turns out that the spinor is already a naturally
4-vector-like quantity, to which Lorentz transformations can be applied.

We will adopt the font $\ffA,\,\ffB,\, \ldots$ for 4-vectors, and use
index notation where convenient. The inner product of 4-vectors
is written either $\ffA \cdot \ffB$ or $\ffA^\mu \ffB_\mu$.  The Minkowski
metric is taken with signature $(-1,1,1,1)$. Note, this
is a widely used convention, but it is not the convention often
adopted in particle physics where $(1,-1,-1,-1)$ is more common.

Let $\bis$ be some arbitrary 1st rank spinor. Under a change of inertial reference frame it will
transform as 
\be
\bis' = \Lambda \bis                \label{spams}
\ee
where $\Lambda$ is a $2 \times 2$ matrix to be discovered. To this end, form the outer product
\be
\bis \bis^\dagger =  \twovec{a}{b} \left( a^*, \, b^* \right)
= \twomat |a|^2 & a b^* \\ b a^* & |b|^2 \etwomat.          \label{uudagger}
\ee
This is (an example of) a 2nd rank spinor, and by definition it must transform as
$\bis \bis^\dagger \rightarrow \Lambda \bis \bis^\dagger \Lambda^\dagger$. 
2nd rank spinors (of the standard, contravariant type)
are defined more generally as objects which transform in this way, i.e.
$X \rightarrow \Lambda X \Lambda^\dagger$.

Notice that the matrix in \eq{uudagger} is Hermitian. Thus outer products of 1st rank
spinors form a subset of the set of Hermitian $2 \times 2$ matrices. 
We shall show that the complete set of Hermitian $2 \times 2$ matrices can be used to represent
2nd rank spinors. 

An arbitrary Hermitian $2 \times 2$ matrix can be written
\be
X = \twomat t+z & x - i y \\ x + i y & t - z \etwomat
  = t I + x \sigma_x + y \sigma_y + z \sigma_z,   \label{XfromIxyz}
\ee
which can also be written
\[
X = \sum_{\mu} {\ffX}^{\mu} \sigma^{\mu}
\]
where we introduced $\sigma^0 \equiv I$.
The summation here is written explicitly, because this is not a tensor
expression, it is a way of creating one sort of object (a 2nd rank spinor)
from another sort of object (a 4-vector).

Evaluating the determinant, we find
\[
|X| = t^2-(x^2+y^2+z^2),
\]
which is the Lorentz invariant associated with the 4-vector $\ffX^\mu$.
Consider the transformation
  \be   X \rightarrow \Lambda X \Lambda^{\dagger}.  \label{MXMdag}
  \ee
To keep the determinant unchanged we must have
\[
|\Lambda| |\Lambda^{\dagger}| = 1  \gap{1} \Rightarrow \;\; |\Lambda| = e^{i\lambda}
\]
for some real number $\lambda$. Let us first restrict attention to $\lambda = 0$. 
Then we are considering complex matrices $\Lambda$ with determinant 1, i.e. the
group SL(2,C). Since the action of members of SL(2,C) preserves
the Lorentz invariant quantity, we can associate a 4-vector
$(t,{\bf r})$ with the matrix $X$, and we can associate a Lorentz
transformation with any member of SL(2,C).

The more general case $\lambda \ne 0$ can be included by considering transformations of
the form $e^{i\lambda/2} \Lambda$ where $|\Lambda| = 1$. It is seen that the additional phase
factor has no effect on the 4-vector obtained from any given spinor, but it rotates the
flag through the angle $\lambda$. This is an example of the fact that spinors are richer
than 4-vectors. However, just as we did not include such global phase factors in our definition
of `rotation', we shall also not include it in our definition of `Lorentz transformation'.
In other words, the group of Lorentz transformation of spinors is the group 
of $2 \times 2$ complex matrices with determinant 1 (called SL(2,C)).

The extra parameter (allowing us to go from a 3-vector to a 4-vector)
compared to eq. (\ref{Xrdotsig}) is exhibited in the $tI$ term. The resulting
matrix is still Hermitian but it no longer needs to have zero trace,
and indeed the trace is not zero when $t \ne 0$.
Now that we don't require the trace of $X$ to be fixed, we can allow
non-unitary matrices to act on it.
In particular, consider the matrix
\be
e^{-(\rho/2) \sigma_z} &=& \twomat e^{-\rho/2} & 0 \\ 0 & e^{\rho/2} \etwomat \nonumber \\
&=& \cosh(\rho/2) I - \sinh(\rho/2) \sigma_z.
    \label{zboostspin}
\ee
One finds that the effect on $X$ is such that the associated 4-vector is
transformed as
\[
\fourmat
\cosh(\rho) & 0 & 0 & -\sinh(\rho) \\
          0 & 1 & 0 & 0 \\
          0 & 0 & 1 & 0 \\
-\sinh(\rho) & 0 & 0 & \cosh(\rho).  \efourmat
\]
This is a Lorentz boost along $z$, with rapidity $\rho$. You can check that
$\exp(-(\rho/2) \sigma_x)$ and $\exp(-(\rho/2) \sigma_y)$ give Lorentz boosts
along $x$ and $y$ respectively. (This must be the case, since the Pauli matrices can
be related to one another by rotations). The general Lorentz boost
for a spinor is, therefore, (for $\boldrho = \rho {\bf n}$)
\be
e^{-(\boldrho/2) \cdot \boldsig} = \cosh(\rho/2) I - \sinh(\rho/2) {\bf n}\cdot \boldsig.
    \label{boostspin}
\ee

We thus find the whole of the structure of the restricted Lorentz group
reproduced in the group SL(2,C). The relationship
is a two-to-one mapping since a given Lorentz transformation
(in the general sense, including rotations) can be
represented by either $+M$ or $-M$, for $M \in $ SL(2,C). The abstract space
associated with the group SL(2,C) has three complex dimensions and therefore
six real ones (the matrices have four complex numbers and one complex constraint
on the determinant). This matches the 6 dimensions of the
manifold associated with the Lorentz group.

Now let
\be
\epsilon = \twomat 0 & 1 \\ -1 & 0 \etwomat.       \label{epsmetric}
\ee
For an arbitrary Lorentz transformation
\[
\Lambda = \twomat a & b \\ c & d \etwomat, \gap{1} ad-bc = 1
\]
we have
\be
\Lambda^T \epsilon \Lambda =
\twomat a & c \\ b & d \etwomat \twomat c & d \\ -a & -b \etwomat = \epsilon
          \label{LepsL}
\ee
It follows that for a pair of spinors $\bis$, $\biw$ the scalar quantity
\[
 \bis^T \epsilon \biw = s_1 w_2 - s_2 w_1
\]
is Lorentz-invariant. Hence this is a useful inner product for spinors.

Equation (\ref{LepsL})
should remind you of the defining property of Lorentz transformations
applied to tensors, ``$\Lambda^T g \Lambda = g$" where $g$ is the Minkowski
metric tensor.
The matrix $\epsilon$ satisfying (\ref{LepsL})
is called the {\bf spinor Minkowski metric}.

A full exploration of the symmetries of spinors involves the recognition that the
correct group to describe the symmetries of
particles is not the Lorentz group but the Poincar\'e group. We shall not explore
that here, but we remark that in such a study the concept of intrinsic spin
emerges naturally, when one asks for a complete set of quantities that can
be used to describe symmetries of a particle. One such quantity is the
scalar invariant $\ffP \cdot \ffP$, which can be
recognised as the (square of the) mass of a particle. A second quantity emerges,
related to rotations, and its associated invariant is
$\ffW \cdot \ffW$ where $\ffW$ is the Pauli-Lubanski spin vector.

\subsection{Obtaining 4-vectors from spinors}      \label{s.4vecspin}

By interpreting \eq{uudagger} using the general form \eq{XfromIxyz} we find that
the four-vector associated with the 2nd rank spinor obtained from the
1st rank spinor $\bis$ is
\be
\fourvec{t}{x}{y}{z}
= \fourvec{(|a|^2 + |b|^2)/2}{(a b^* + b a^*)/2}{i(a b^* - b a^*)/2}{(|a|^2 - |b|^2)/2}
&=& \frac{1}{2} \twovec{\bra{s} I \ket{s}}{\bra{s} \boldsig \ket{s}}
             \label{u_to_4vec}
\ee
which can be written $(1/2) \bra{s} \sigma^{\mu} \ket{s}$.
Any constant multiple of this is also a legitimate 4-vector. In order that the spatial
part agrees with our starting point \eq{def_ab} we must introduce\footnote{When 
moving between the 4-vector and the complex number representation, the overall scale
factor is a matter of convention. The convention adopted here slightly simplifies various
results. Another possible convention is to retain the
factor $1/2$ as in \protect\eq{u_to_4vec}.} a factor 2, so
that we have the result (perhaps the central result of this introduction)

\eqbox
{obtaining a (null) 4-vector from a spinor}
{\ffV^{\mu} &=& \biv^{\dagger} \sigma^{\mu} \biv}
{get_flagpole}
\vspace{6pt}\\
This 4-vector is null, as we mentioned
in the introductory section \ref{s.introspinor}. The easiest way to verify this
is to calculate the determinant of the spinor matrix (\ref{uudagger}). 

Since the zeroth spin matrix is the identity, we find that the zeroth
component of the 4-vector can be written $\biv^{\dagger} \biv$. This and
some other basic quantities are listed in table~\ref{t.spin_scalar}.

\begin{table}
\begin{tabular}{lll}
$\biu^{\dagger} \biu$ & zeroth component of a 4-vector \\
$\biu^T \epsilon \biu$ & a scalar invariant (equal to zero) \\
$\overline\biu^{\dagger} \biu$ & another way of writing $\biu^T \epsilon \biu$,
with $\overline\biu \equiv \epsilon \biu^*$ \\
$\biu^T \biu$ & no particular significance
\end{tabular}
\caption{Some scalars associated with a spinor, and their significance.}
\label{t.spin_scalar}
\end{table}

The linearity of eq. (\ref{spams}) shows that the sum of two spinors is also a spinor
(i.e. it transforms in the right way). The new spinor still corresponds to
a null 4-vector, so it is in the light cone. Note, however, that the sum of two
null 4-vectors is not in general null. So adding up two spinors as in $\biw =
\biu + \biv$ does not result in a 4-vector $\ffW$
that is the sum of the 4-vectors $\ffU$ and $\ffV$ associated with each of the spinors.
If you want to get access to $\ffU + \ffV$, it is easy to do:
first form the outer product, then sum:
$\biu \biu^{\dagger} + \biv \biv^{\dagger}.$ The resulting $2 \times 2$ matrix
represents the (usually non-null) 4-vector $\ffU + \ffV$.

\begin{figure}
\myfig{0.3}{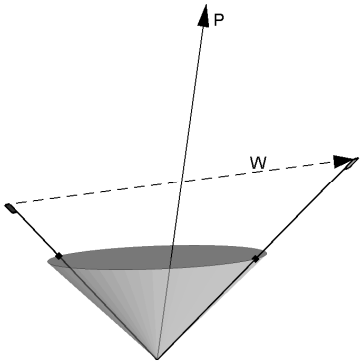}
\caption{Two spinors can represent a pair of orthogonal 4-vectors.
The spacetime diagram shows two spinors. They have opposite spatial direction and
are embedded in a null cone (light cone), including the flags which point around
the cone. Their amplitudes are not necessarily equal.
The sum of their flagpoles is a time-like 4-vector $\ffP$; the
difference is a space-like 4-vector $\ffW$. $\ffP$ and $\ffW$ are orthogonal
(on a spacetime diagram this orthogonality is shown by the fact that
if $\ffP$ is along the time axis of some reference frame, then $\ffW$ is
in along the corresponding space axis.)}
\label{f.twospinor}
\end{figure}

By using a pair of non-orthogonal null spinors, we can always represent a pair
of orthogonal non-null 4-vectors by combining the spinors.
Let the spinors be $\biu$ and $\biv$ and their associated 4-vectors be $\ffU$ and $\ffV$.
Let $\ffP = \ffU + \ffV$ and $\ffW = \ffU - \ffV$.
Then $\ffU\cdot\ffU = 0$ and $\ffV \cdot \ffV = 0$ but $\ffP \cdot \ffP = 2 \ffU \cdot \ffV \ne 0$
and $\ffW \cdot \ffW = -2 \ffU \cdot \ffV \ne 0$. That is, as long as $\ffU$ and $\ffV$ are not
orthogonal then $\ffP$ and $\ffW$ are not null. The latter are orthogonal to one another, however:
\[
\ffW \cdot \ffP = (\ffU + \ffV) \cdot (\ffU - \ffV) = \ffU \cdot \ffU - \ffV \cdot \ffV = 0.
\]
Examples of pairs of 4-vectors that are mutually orthogonal are 4-velocity and 4-acceleration,
and 4-momentum and 4-spin (i.e. Pauli-Lubanski spin vector \cite{12Steane}).
Therefore we can describe the motion and spin of a particle by using a pair of spinors,
see figure \ref{f.twospinor}.
This connection will be explored further in section \ref{s.spinorapp}.

To summarize:\\

\fbox{
  \begin{minipage}[h]{0.41\textwidth}
\begin{tabular}{lcl}
rank 1 spinor & $\leftrightarrow$ & null 4-vector \\
rank 2 spinor & $\leftrightarrow$ & arbitrary 4-vector\\
\begin{tabular}{l}
pair of non-orthogonal \\
rank 1 spinors \end{tabular} & $\leftrightarrow$ & \begin{tabular}{l}pair of orthogonal \\4-vectors\end{tabular}
\end{tabular}
  \end{minipage}
}

\section{Chirality}  \label{s.chirality}

We now come to the subject of {\em chirality}. This concerns a property of
spinors very much like the property of {\em contravariant} and {\em covariant}
applied to 4-vectors. In
other words, chirality is essentially about {\em the way spinors transform under
Lorentz-transformations}. Unfortunately, the name itself does not suggest that.
It is a bad name. In order to understand this we shall discuss the transformation
properties first, and then return to the terminology at the end.

First, let us notice that there is another way to construct a contravariant
4-vector from a spinor. Suppose that instead of
(\ref{get_flagpole}) we try
\be
\ffV^{\mu} &=&  \twovec{\bra{\tilde{v}} -I \ket{\tilde{v}}}
{\bra{\tilde{v}} \boldsig \ket{\tilde{v}}}
=  \bra{\tilde{v}} \sigma_{\mu} \ket{\tilde{v}},    \label{guess_tildeu}
\ee
for a spinor-like object $\tilde\biv$. It looks at first as
though we have constructed a covariant
4-vector and put the index `upstairs' by mistake. However, what if we insist that
this $\ffV$ really is contravariant? This amounts to saying that $\tilde\biv$ is a new
type of object, not like the spinors we talked about up till now. To explore this,
observe that the same assignment can also be written
\be
\ffV_{\mu} =  \bra{\tilde{v}} \sigma^{\mu} \ket{\tilde{v}}.     \label{Vfrom_tildeu}
\ee
Index notation does not lend itself to the proof that (\ref{Vfrom_tildeu}) and (\ref{guess_tildeu})
imply each other, but it can be seen readily enough by using a rectangular coordinate
system and writing out all the terms, since there the Minkowski metric has the simple form
$g_{ab} = {\rm diag}(-1,1,1,1)$ and its inverse is the same, $g^{ab}={\rm diag}(-1,1,1,1)$.
We deduce that the difference between $\biv$ and $\tilde\biv$ is that when combined
with $\sigma^{\mu}$, the former gives a contravariant and the latter gives
a covariant 4-vector. Everything is consistent if we
introduce the rule for a Lorentz transformation of $\tilde\biv$ as
\be
\mbox{if } \; \biv' &=& \Lambda \biv \\
\mbox{then } \;  \tilde\biv' &=& (\Lambda^{\dagger})^{-1} \tilde\biv.
\ee
This is because, for a pure rotation $\Lambda^{\dagger} = \Lambda^{-1}$
so the two types of spinor transform the {\em same} way, but for a pure boost
$\Lambda^{\dagger} = \Lambda$ (it is Hermitian) so we have precisely the
inverse transformation. This combination of properties is exactly
the relationship between covariant and contravariant 4-vectors.

The two types of spinor may be called {\em contraspinor} and {\em cospinor}.
However, they are often called {\em right-handed} and {\em left-handed}. The idea
is that we regard the Lorentz boost as a kind of `rotation in spacetime', and
for a {\em given} boost velocity, the contraspinor `rotates' one way, while
the cospinor `rotates' the other. They are said to possess opposite
{\em chirality}. However, given that we are also much concerned with real
rotations in space, this terminology is regretable because it leads to
confusion.

Equation (\ref{Vfrom_tildeu}) can be `read' as stating that the presence of
$\tilde\biv$ acts to lower the index on $\sigma^{\mu}$ and give a covariant
result.

The rule (\ref{guess_tildeu}) was here introduced ad-hoc: what is to say
there may not be further rules? This will be explored below; ultimately the quickest
way to show this and other properties is to use Lie group theory on the generators,
a method we have been avoiding in order not to assume familiarity with groups, but it is
briefly sketched in section (\ref{s.spin_algebra}).

\subsection{Chirality, spin and parity violation}   \label{s.chiral_spin}

It is not too surprising to suggest that a spinor may offer a useful mathematical tool to
handle angular momentum. This was the context in which spinors were first widely
used. A natural way to proceed is simply to claim that there may exist fundamental
particles whose intrinsic nature is not captured purely by scalar properties
such as mass and charge, but which also have an angular-momentum-like property
called spin, that is described by a spinor.

Having made the claim, we might propose that the 4-vector represented by the
spinor flagpole is the Pauli-Lubanski spin vector \cite{12Steane}. The Pauli-Lubanski vector
has components
\be
\ffW^\mu = \left(\vs \cdot \vp, (E/c) \vs \right)
\ee
for a particle with spin 3-vector $\vs$, energy $E$ and momentum $\vp$. If
this 4-vector can be extracted from a rank 1 spinor, then it must be a null 4-vector.
This in turn implies the particle is massless, because
\be
\ffW^\mu \ffW_\mu =0 \implies E^2 = p^2 c^2 \cos^2 \theta
\label{spintheta}
\ee
where $\theta$ is the angle between $\vs$ and $\vp$ in some reference frame.
But, for any particle, $E \ge pc$, so the only possibility is $E=pc$ and
$\theta=0$ or $\theta=\pi$.
Therefore we look for a massless spin-half particle in our experiments. We already
know one: it is the neutrino\footnote{There now exists strong evidence that neutrinos
possess a small non-zero mass. We proceed with the massless model as a valid theoretical device,
which can also serve as a first approximation to the behaviour of neutrinos.}.

Thus we have a suitable model for intrinsic spin, that applies to massless
spin-half particles. It is found in practice that it describes accurately the experimental
observations of the nature of intrinsic angular momentum for such particles.

Now we shall, by `waving a magic wand', discover a wonderful property
of massless spin-half particles that emerges naturally when we use spinors, but
does not emerge naturally in a purely 4-vector treatment
of angular momentum (as, for example, in chapter 15 of \cite{12Steane}).
By `waving a magic wand' here we mean noticing something that
is already built in to the mathematical properties of the objects we are dealing with,
namely spinors. All we need to do is claim that the same spinor describes both the
linear momentum and the intrinsic spin of a given neutrino. We claim that we
don't need two spinors to do the job: just one is sufficient.
There is a problem: since we can only allow one rule for extracting the 4-momentum and
Pauli-Lubanski spin vector for a given type of particle, we shall have to claim that
there is a restriction on the allowed combinations of 4-momentum and spin for
particles of a given type. For massless particles the Pauli-Lubanski spin
and the 4-momentum are aligned (either in the same direction or opposite directions, 
as we showed after eqn (\ref{spintheta})), so there is already one restriction that 
emerges in either a 4-vector or a spinor analysis,
but now we shall have to go further, and claim that
{\em all massless spin-half particles of a given type have the same helicity}
(equations (\ref{poshel}) and (\ref{neghel})).

This is a remarkable claim, at first sight even a crazy claim. It says that,
relative to their direction of motion,
neutrinos are allowed to `rotate' one way, but not the other! To be more precise,
it is the claim that there exist in Nature processes whose mirror reflected
versions never occur.
Before any experimenter would invest the effort to test this (it is difficult to test because neutrinos interact
very weakly with other things), he or she would want more convincing
of the theoretical background, so let us investigate further.

Processes whose mirror-reflected versions run differently
(for example, not at all) are said to exhibit {\bf parity violation}.
We can prove that there are no such processes in classical electromagnetism,
because Maxwell's equations and the Lorentz force equation
are unchanged under the parity inversion operation.
The `parity-invariant' behaviour of the last two Maxwell equations, and the Lorentz
force equation, involves the fact that $\vB$ is an axial vector.

To investigate the possibilities for spinors, consider the Lorentz invariant
\[
\ffW_\lambda \ffS^\lambda = \ffW_\lambda \bis^\dagger \sigma^\lambda \bis
\]
where $\bis$ is contravariant. Since, in the sum, each term $\ffW_\lambda$ is just a number,
it can be moved past the $\bis^\dagger$ and we have
\be
\ffW_\lambda \ffS^\lambda = \bis^\dagger \ffW_\lambda \sigma^\lambda \bis.
\label{Wsigs}
\ee
The combination $\ffW_\lambda \sigma^\lambda = -\ffW^0 I + \vw \cdot \sigma$ is a matrix.
It can usefully be regarded as an operator acting on a spinor. We can prove that one
effect of this kind of matrix, when multiplying a spinor, is to change
the transformation properties. For, $\bis$ transforms as 
\[
\bis \rightarrow \Lambda \bis \gap{1}
\]
and therefore
\[
\bis^{\dagger} \rightarrow \bis^{\dagger} \Lambda^{\dagger}.
\]
Since $\ffW_\lambda \ffS^\lambda$ is invariant, we deduce from (\ref{Wsigs})
that $\ffW_\lambda \sigma^\lambda \bis$
must transform as
\be
(\ffW_\lambda \sigma^\lambda \bis) \rightarrow (\Lambda^{\dagger})^{-1} 
(\ffW_\lambda \sigma^\lambda \bis).
    \label{convert_cc}
\ee
Therefore, for any $\ffW$, if $\bis$ is a contraspinor then
$(\ffW_\lambda \sigma^\lambda \bis)$ is a cospinor, and {\em vice versa}.

\miniexplainbox{{\bf Notation}. We now have 3 vector-like quantities in play: 3-vectors,
4-vectors, and rank 1 spinors. We adopt three fonts:

\begin{tabular}{lll}
entity & font & examples \\
3-vector & bold upright Roman & $\vs, \; \vu, \; \vv, \; \vw$ \\
4-vector & sans-serif capital & $\ffS, \; \ffU, \; \ffV, \; \ffW$ \\
spinor & bold italic & $\bis, \; \biu, \; \biv, \; \biw$
\end{tabular}
}

If the 4-vector $\ffW$ is null, then it can itself be represented by a spinor $\biw$. Let's
see what happens when the matrix $\ffW_\lambda \sigma^\lambda$ multiplies the spinor
representing $\ffW$:
\be
\ffW_\lambda \sigma^\lambda  \biw
&=&
\twomat
-2|b|^2 & 2 a b^* \\
2 a^* b & -2|a|^2
\etwomat
\twovec{a}{b}    = \twovec{0}{0}                                      \label{Wsigu}
\ee
where for convenience we worked in terms of the components 
($\biw = (a,b)^T$) in some reference frame. The result is
\be
(-\ffW^0 + {\bf w} \cdot \boldsig) \biw = 0   \label{Wdotsig}
\ee
(N.B. in this equation $\vw$ is a 3-vector whereas $\biw$ is a spinor).
This equation is important because it is (by construction) a Lorentz-covariant
equation, and it tells us something useful about 1st rank spinors in general.

Suppose the 4-vector $\ffW$ is the 4-momentum of some massless particle. Then
the equation reads
\be
(E/c - {\bf p} \cdot \boldsig) \biw = 0.   \label{Weyl_v1}
\ee
This equation is called the first {\bf Weyl equation} and in the context of
particle physics, the rank 1 spinors are called {\em Weyl spinors}. The presence
of $\boldsig$ in this equation invites us to guess that the equation might be
interpreted also as a statement about intrinsic spin. This guess is very natural
if we suppose that a single spinor can serve to encode both the
4-momentum and the 4-spin, for massless spin-half particles, and this is the
interpretation proposed by Weyl. To adopt this interpretation, it is necessary to use a polar
version of the vector $\boldsig$ when using (\ref{get_flagpole}) to relate
$\biw$ to linear momentum, and an axial version
to extract the spin (which, being a form of angular momentum, must be axial).
Therefore when the Weyl equation \eq{Weyl_v1} is used in this context, $\vp$ is polar but
$\boldsig$ is axial. This means the equation transforms in a non-trivial way under
parity inversion. In short, it is not parity-invariant. This was enough to make particle
physicists very dubious of the claim that the equation could describe real
physical behaviour---but it turns out that Nature does admit this type
of behaviour, and neutrinos give an example of it.

For a massless particle, we have $E=pc$, so (\ref{Weyl_v1}) gives
\be
\frac{({\bf p} \cdot \boldsig)}{p} \biw = \biw.      \label{poshel}
\ee
This says that $\biw$ is an eigenvector, with eigenvalue 1, of the spin operator
pointing along $\bf p$. In other words, {\em the particle has positive helicity}.

Now let's explore another possibility: suppose the spinor representing the
particle has the other chirality. Then the energy-momentum is obtained as
\be
\ffP_{\mu} &=& \tilde\biv^{\dagger} \sigma^{\mu} \tilde\biv.
\ee
where the use of a different letter ($\biv$) indicates that we are talking
about a different particle, and the tilde acts as a reminder of the different
transformation properties. The invariant is now
\be
\ffP^\lambda \ffS_\lambda = 
\tilde{\bis}^\dagger \ffP^\lambda \sigma^\lambda \tilde{\bis}
= \tilde{\bis}^\dagger (\tilde{\biv}^\dagger \sigma_\lambda \tilde{\biv}) \sigma^\lambda \tilde{\bis}
\ee
and the operator of interest is
\be
\left( \tilde\biv^{\dagger} \sigma_\lambda \tilde\biv \right) \sigma^\lambda
= E/c + {\bf p} \cdot \boldsig.
\ee
The version on the right hand side does not at first sight look like a Lorentz invariant, because
of the absence of a minus sign, but as long as we use the operator with cospinors (left handed spinors)
then Lorentz covariant equations will result. For example,
the argument in (\ref{Wsigu}) is essentially unchanged and we find
\be
\left( \tilde\biv^{\dagger} \sigma^{\alpha} \tilde\biv \right)
\sigma_{\alpha} \tilde\biv &=& 0   \\
\mbox{i.e. } \gap{1}
 (E/c + {\bf p} \cdot \boldsig)\tilde\biv &=& 0.      \label{Weyl_v2}
\ee
This is called the 2nd Weyl equation. Since the particle is massless it implies
\be
\frac{({\bf p} \cdot \boldsig)}{p} \tilde\biv &=& - \tilde\biv.    \label{neghel}
\ee
Therefore now the helicity is negative.

\begin{figure}
\myfig{0.2}{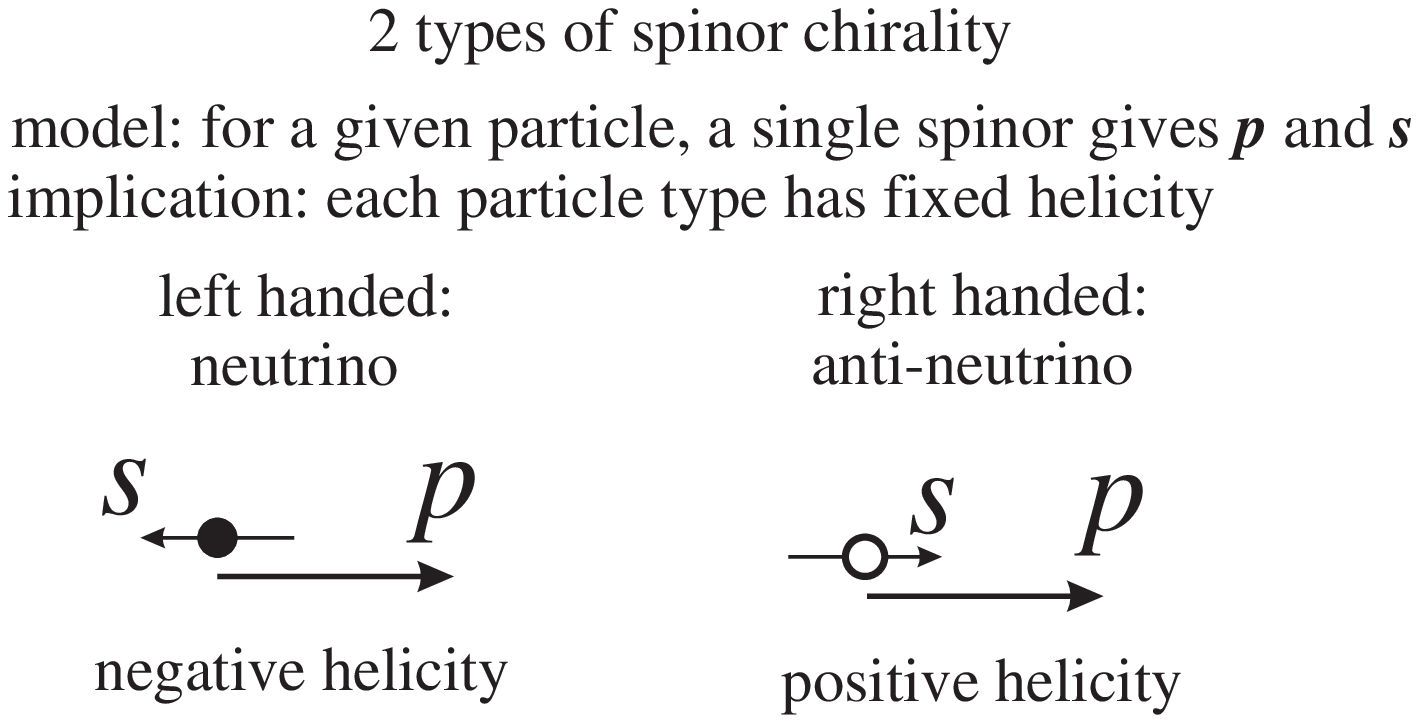}
\caption{The black and white circles represent two particle-like entities. Both are massless
leptons with spin $1/2$ and zero charge. Are they two examples of the same type
of particle then, merely having the spin in opposite directions? The sequence of statements
shown in the figure gives the logic. The black entity is found to have different chirality
from the white entity. This is a subtle property, not easily illustrated by any diagram,
since it refers to how the spinor transforms under a boost. However this property suffices
to distinguish one entity from the other, and it is legitimate to give them different
names (``neutrino" and ``anti-neutrino") and draw them with different colours. The theoretical
model asserts that the information about 4-spin and energy-momentum is contained in
a single spinor for each entity. It then follows that the helicity is single-valued: always
negative for the one we called ``neutrino" and always positive for the one we 
called ``anti-neutrino".
Similar reasoning applied to electrons reaches a different conclusion. Each electron is not described
by a single spinor, but by a pair of spinors, one of each chirality. Consequently the helicity
of an electron can be of either sign, and is not Lorentz-invariant.}
\end{figure}

\explainbox
{{\bf What is the difference between chirality and helicity?}\\
Answer: helicity refers to the projection of the spin along the direction of motion,
chirality refers to the way the spinor transforms under Lorentz transformations.\\

The word `chirality' in general in science refers to
{\em handedness}. A screw, a hand, and certain types of molecule may be said to
possess {\em chirality}. This means they can be said to embody a rotation that is
either left-handed or right-handed with respect to a direction also embodied by
the object. When Weyl spinors are used to represent spin angular momentum and
linear momentum, they also possess a handedness, which can with perfect sense be
called an example of chirality. However, since the particle physicists already
had a name for this (helicity), the word chirality came to be used
to refer directly to the transformation property, such that
spinors transforming one way are said to be `right-handed'
or of `positive chirality', and those transforming the other way
are said to be `left-handed' or of `negative chirality.'
This terminology is poor because (i) it invites (and in practice results in)
confusion between chirality and helicity, (ii) spinors can be used to describe
other things beside spin, and (iii) the {\em transformation rule} has nothing
in itself to do with angular momentum. The terminology is acceptable, however,
if one understands it to refer to the Lorentz boost as a form of `rotation' in
spacetime.}

Overall, the spinor formalism suggests that there are two particle types,
possibly related to one another in some way, but they are not
interchangeable because they transform in different ways under Lorentz
transformations. We are then forced to the `parity-breaking' conclusion
that one of these types of particle always has
positive helicity, the other negative. This is born out in experiments.
An experimental test involving the $\beta$-decay of cobalt nuclei was performed in
1957 by Wu {\em et al.}, giving
clear evidence for parity non-conservation. In 1958 Goldhaber {\em et al.} took
things further in a beautiful experiment, designed to allow
the helicity of neutrinos to be determined. It was found that all neutrinos produced
in a given type of process have the same helicity. This is evidence that
all neutrinos have one helicity, and anti-neutrinos have the opposite
helicity. By convention those with positive helicity are
called anti-neutrinos. With this convention, the process
\be
{\rm n} \rightarrow {\rm p} + {\rm e} + \bar{\nu}
\ee
is allowed (with the bar indicating an antiparticle), but the process
${\rm n} \rightarrow {\rm p} + {\rm e} + \nu$ is not. Thus the properties
of Weyl spinors are at the heart of the parity-non-conservation exhibited
by the weak interaction.

\subsection{Reflection and Lorentz transformation}
\ifodd\final

Recall the relationship between a spinor $\bis$ and the 3-vector of its flagpole:
\[
\vr = \bis^\dagger \boldsig \bis.
\]
Taking the complex conjugate yields
\[
\vr^* = \vr = (\bis^*)^\dagger \boldsig^* \bis^*.
\]
Now, since $\sigma_x$ and $\sigma_z$ are real while $\sigma_y$ is imaginary, 
$\boldsig^* = (\sigma_x, \; -\sigma_y, \; \sigma_z)$. Therefore
\[
(\bis^*)^\dagger \boldsig \bis^* = (x,\;-y,\;z).
\]
In other words, taking the complex conjugate of a spinor corresponds to a reflection
in the $xz$ plane. An inversion through the origin (parity inversion) is obtained
by such a reflection followed by a rotation about the $y$ axis through 180\degree,
i.e. the transformation
\be
\bis \rightarrow e^{i (\pi/2) \sigma_y} \bis^* = \twomat 0&1\\-1&0\etwomat \bis^* = \epsilon \bis^*
\ee
where the last version uses the spinor 
Minkowski metric $\epsilon$ introduced in eq. (\ref{epsmetric}).

We now have four possibilities: for a given $\bis$ we can construct
three others by use of complex conjugation
and multiplication by the metric. These are $\bis^*$, $\epsilon \bis$ and
$\epsilon \bis^*$.
They transform under a general Lorentz transformation as:
\be
\bis &\rightarrow& \Lambda \bis  \nonumber  \\
\bis^* &\rightarrow& \Lambda^* \bis^*    \nonumber  \\
(\epsilon \bis) & \rightarrow &  (\Lambda^T)^{-1} (\epsilon \bis)   \nonumber   \\
(\epsilon \bis^*) & \rightarrow &  (\Lambda^{\dagger})^{-1} (\epsilon \bis^*)
             \label{chirality}
\ee
The second result follows immediately from the first by complex conjugation. The third
result uses $\epsilon \Lambda = (\Lambda^T)^{-1}\epsilon$ from (\ref{LepsL}); the
fourth then follows by complex conjugation. The last result shows that parity
inversion changes the chirality. That is, under parity inversion,
a right handed spinor changes into a left handed one, and {\em vice versa}.

A pure boost such as $\exp(-(\rho/2) \sigma_z)$ is Hermitian. From (\ref{zboostspin})
we have
\[
e^{(-\rho/2) \sigma_z} = \cosh(\rho/2) I - \sinh(\rho/2) \sigma_z
\]
and therefore
\be
\left(e^{(-\rho/2) \sigma_z}\right)^{-1} = e^{(\rho/2) \sigma_z} .
\ee
This confirms, as expected, that the inverse Lorentz boost is obtained by reversing the
sign of the velocity. Thus we deduce that $\epsilon \bis^*$
transforms the same way as $\bis$ under rotations, but it
transforms the inverse way (i.e. with opposite velocity sign) under boosts. This confirms
that it is the covariant partner to $\bis$. Sometimes $\epsilon \bis^*$ is called
the {\bf dual} of $\bis$ and is written $\overline\bis \equiv \epsilon \bis^*$.

\begin{table*}
\begin{tabular}{c|rrrc}
spinor & any transformation & pure rotation & pure boost & chirality \\
\hline
$\biu$ &                          $\Lambda \biu \gap{1}$ & $U \biu\gap{.6}$ & $L \biu$ & $+$ \\
$\biv = (\epsilon \biu^*)$ & $(\Lambda^{\dagger})^{-1} \biv\gap{1}$  & $U \biv\gap{.6}$ & $L^{-1} \biv$ & $-$ \\
\hline
$\bis = \biu^*$           & $\Lambda^{\!*}\, \bis\gap{1}$ & $U^* \, \bis\gap{.6}$ & $ L^T \, \bis$ & $+$ \\
$\bit = (\epsilon \biu)$ & $(\Lambda^{-1})^T \bit\gap{1}$  & $U^* \bit\gap{.6}$ & $(L^{-1})^T \bit$ & $-$
\end{tabular}
\caption{Four types of spinor and their transformation. In principle, any expression can be written
using just one of these types of spinor, by including explicit use of $\epsilon$ and complex
conjugation (see exercise 4). In practice, a notation such as $\biu$ and $\tilde\biv$ or
$\phi_R$, $\chi_L$ is more convenient to write the first two types, then complex
conjugation suffices to express the other two types where needed.}
\label{spin.t.trans}
\end{table*}

The results are summarized in table \ref{spin.t.trans}.
The Lorentz transformation for the $\epsilon s^*$ case can also be written
\be
 (\Lambda^{\dagger})^{-1} = \epsilon \Lambda^* \epsilon^{-1}         \label{Lamdagger}
\ee
(this is quickly proved for arbitrary $\Lambda \in $ SL(2,C) using (\ref{epsmetric}).)


In short, we have discovered that there are four types of spinor, distinguished by how
they behave under a change of inertial reference frame. These are best described as
two types, plus their mirror inversions:
\begin{enumerate}
\item Type I, called `right handed spinor' or `positive chirality spinor'
\[
\phi_R   \rightarrow  \Lambda \phi_R  =
\exp\left(i \frac{\boldsig \cdot \boldtheta}{2} - \frac{\boldsig \cdot {\boldrho}}{2} \right)
\phi_R
\]
\item Type II, called `left handed spinor' or `negative chirality spinor'
\[
\phi_L    \rightarrow    (\Lambda^{\dagger})^{-1} \phi_L   =
\exp\left(i \frac{\boldsig \cdot \boldtheta}{2} + \frac{\boldsig \cdot {\boldrho}}{2} \right)
\phi_L
\]
\end{enumerate}



\else\omission\fi

\subsection{Index notation*}
\ifodd\final

Suppose we take a 2nd rank spinor $X$ obtained from a four-vector following the prescription
in eq. (\ref{XfromIxyz}), and a first rank spinor $\biu$ transforming as $\biu \rightarrow \Lambda \biu$.
We might be tempted to evaluate the product $X \biu$ (i.e. a $2 \times 2$ matrix multiplying a
column vector), but we must immediately check whether or not the result is a spinor. It is not.
Proof: $X \rightarrow \Lambda X \Lambda^{\dagger}$ and $\biu \rightarrow \Lambda \biu$ so
$X \biu \rightarrow \Lambda X \Lambda^{\dagger} \Lambda \biu$ which is not equal to
$\Lambda X \biu$ nor does it correspond to
any of the other transformations listed in table \ref{spin.t.trans}.

A similar issue arises with 4-vectors and tensors, and it is handled by involving the metric tensor $g$.
The index notation signals the presence of $g$ by a lowered index; the matrix notation signals
the presence of $g$ by the dot notation for an inner product.
When using index notation, a contraction is only a legal tensor operation
if it involves a pair consisting of one upper and one lower index.
For spinor manipulations, a similar notation is available, but we have a further complication:
there are four kinds of basic spinor, not just two. This leads to 16 kinds of 2nd rank spinors, 64 kinds
of 3rd rank spinors, and so on. Fortunately, just as with tensors, the higher rank spinors all transform
in the same way as outer products of lower rank spinors, so the whole system can be `tamed' by the
use of index notation.

\begin{table*}
\[
\begin{array}{r|cc|cc}
  M^{\bullet \bullet}       & v^{\bullet} & v_{\bullet} & v^{\star} & v_{\star} \\
 = \biu \biv^T & \biv & (\epsilon \biv) & v^* & (\epsilon \biv^*) \\
\hline
u^{\bullet} = \biu & \Lambda M^{\bullet\bullet} \Lambda^T & \Lambda M^{\bullet}_{\;\, \bullet} \Lambda^{-1} &
  \Lambda M^{\bullet\star} \Lambda^{\dagger} & \Lambda M^{\bullet}_{\;\,\star} (\Lambda^*)^{-1}  \\
u_{\bullet} = \epsilon \biu & (\Lambda^T)^{-1} M_{\bullet}^{\;\,\bullet} \Lambda^T
  & (\Lambda^T)^{-1} M_{\bullet\bullet} \Lambda^{-1} &
  (\Lambda^T)^{-1} M_{\bullet}^{\;\,\star} \Lambda^{\dagger} & (\Lambda^T)^{-1} M_{\bullet\star} (\Lambda^*)^{-1}  \\
\hline
u^{\star} = \biu^* & \Lambda^* M^{\star\bullet} \Lambda^T & \Lambda^* M^{\star}_{\;\, \bullet} \Lambda^{-1} &
  \Lambda^* M^{\star\star} \Lambda^{\dagger} & \Lambda^* M^{\star}_{\;\,\star} (\Lambda^*)^{-1}  \\
u_{\star} = \epsilon \biu^* & (\Lambda^{\dagger})^{-1} M_{\star}^{\;\,\bullet} \Lambda^T
  & (\Lambda^{\dagger})^{-1} M_{\star\bullet} \Lambda^{-1} &
  (\Lambda^{\dagger})^{-1} M_{\star}^{\;\,\star} \Lambda^{\dagger} & (\Lambda^{\dagger})^{-1} M_{\star\star} (\Lambda^*)^{-1}
\end{array}
\]
\caption{Transformation rules for 2nd rank spinors. The first row and column show the four types of rank-1 spinor.
In the table, the $M$ symbols are 2nd rank spinors formed from the outer product of the rank-1 spinor of each
row and column. For example, $M^{\bullet}_{\;\,\bullet} = \biu (\epsilon \biv)^T$.
The dots and stars attached to $M$ symbols serve as generic indices of one of two types.
The entries in the table show how each
$M$ transforms under a change of reference frame (see text). The table shows, for example, that a matrix product such as
$X^{\bullet}_{\;\,\bullet} Y^{\bullet\star}$ is legal because the transformation carries it to
$\Lambda X^{\bullet}_{\;\,\bullet} \Lambda^{-1} \Lambda Y^{\bullet\star} \Lambda^{\dagger}
=\Lambda X^{\bullet}_{\;\,\bullet} Y^{\bullet\star} \Lambda^{\dagger}$ and furthermore the object that results
is one which transforms as $W^{\bullet\star}$. Thus legal operations and the class of the result
are easily identified by paying attention to the placement and type of index. A spinor of type
$M^{\bullet\star}$ would be written in index notation as $M^{\mu\bar{\nu}}$.}
\label{spin.t.rank2}
\end{table*}

We show in table \ref{spin.t.rank2} all the possible types of 2nd rank spinor,
in order to convey the essential idea. We introduce $\bullet$ and $\star$ symbols
attached to a letter $M$ to serve as a `code' to show what type of spinor is
represented by the matrix $M$.
For example, consider the entry in the first row, second column:
$M^{\bullet}_{\;\,\bullet} = \biu (\epsilon \biv)^T = \biu \biv^T \epsilon^T.$ When $\biu \rightarrow
\Lambda \biu$ and $\biv \rightarrow \Lambda \biv$ we have
\[
M^{\bullet}_{\;\,\bullet} \rightarrow \Lambda \biu \biv^T \Lambda^T \epsilon^T =
\Lambda \biu \biv^T \epsilon^T   \Lambda^{-1} = \Lambda M^{\bullet}_{\;\,\bullet} \Lambda^{-1}
\]
where the second step used the complex conjugate of eq. (\ref{LepsL}), namely
\[
\Lambda^T \epsilon^T \Lambda = \epsilon^T.
\]
By similar arguments you can prove any other entry in the table. In practice one does not need
all the different types of spinor, and we shall be mostly concerned with the types $M^{\bullet\star}$
and $M^{\bullet}_{\;\,\bullet}$. When we go over to index notation, the $\bullet$ will be replaced by a generic
greek letter such as $\mu$, and the $\star$ by a barred letter such as $\bar{\nu}$.

The important result of this analysis is that to ensure we only carry out legal spinor manipulations,
it is sufficient to follow the rule that only indices of the same type can be summed over, and they
must be one up, one down.
That this is true in general, for spinors of all ranks, follows immediately
from (\ref{LepsL}) and its complex conjugate ($\Lambda^{\dagger} \epsilon \Lambda^* = \epsilon$)
as long as we arrange (as we have done) that the index lowering operation is achieved by premultiplying by
$\epsilon$ or postmultiplying by $\epsilon^T$, and
the index raising operation is achieved by premultiplying by $\epsilon^{-1} = \epsilon^T$ or postmultiplying
by $\epsilon$. This applies to either type of index:
\[
M_{\bullet\bullet} = \epsilon M^{\bullet}_{\;\,\bullet}  = M_{\bullet}^{\;\,\bullet} \epsilon^T, \gap{1}
M_{\star\bullet} = \epsilon M^{\star}_{\;\,\bullet},
\]
etc. and
\[
M^{\bullet\bullet} = \epsilon^T M_{\bullet}^{\;\,\bullet}  = M^{\bullet}_{\;\,\bullet} \epsilon, \gap{1}
M^{\star\bullet} = \epsilon^T M_{\star}^{\;\,\bullet},
\]
etc.
The proof that contraction can be applied to spinors of higher rank is
simple: we {\em define} spinors of arbitrary
rank to be entities that transform in the same way as outer products of rank 1 spinors.

We have seen how to raise and lower indices. One may also want to ask, can we convert
a spinor with one type of index to a spinor with another type? For example, can we convert
between $M^{\bullet\bullet}$ and $M^{\bullet\star}$? The answer is that it is not possible
to do this. There is no simple relationship between these two types of spinor. However,
it is possible to change the type of all the indices at once: the matrix $M^{\star\star}$,
for example, is the complex conjugate of the matrix $M^{\bullet\bullet}$, and
$M^{\bullet\star}$ is the complex conjugate of $M^{\star\bullet}$.

Although the Lorentz transformation $\Lambda$ is not itself a spinor (it cannot be written in
any given reference frame, it is a bridge between reference frames), it is convenient
to write it in index notation as $\Lambda^{\mu}_{\;\,\nu}$. Then the transformation of
the standard right-handed spinor can be written $u'^{\mu} = \Lambda^{\mu}_{\;\,\alpha} u^{\alpha}$.
The standard left-handed spinor would be written $v_{\bar{\mu}}$ so its transformation rule should
be $v_{\bar{\mu}}' = \Lambda_{\bar{\mu}}^{\;\,\bar{\alpha}} v_{\bar{\alpha}}$. The relationship
between these two Lorentz transformations is
\be
\Lambda_{\bar{\mu}}^{\;\,\bar{\nu}}
= (\epsilon_{\mu\alpha} \Lambda^{\alpha}_{\;\,\beta} \epsilon^{\beta\nu} )^*.
\ee
This is eq (\ref{Lamdagger}), so everything is consistent (the overall complex conjugation
on the right hand side causes the indices to change from unbarred to barred).

\subsection{Invariants}

The most basic spinor invariant is
\[
u^{\mu} u_{\mu} = 0.      \gap{1} [ = \biu^T \epsilon \biu
\]
That is, the `length' of a spinor, as indicated by this type of scalar product, is zero. This is
consistent with the fact that the flagpole of a spinor is a null 4-vector. To prove the result you
can use $u^{\mu} u_{\mu} = \biu^T \epsilon \biu = u^1 u^2 - u^2 u^1 =0$ or use the general
property that the scalar product of spinors is anticommutative:
\begin{eqnarray*}
u^{\mu} v_{\mu} = u^{\mu} \epsilon_{\mu \alpha} v^{\alpha} &=&
\epsilon_{\mu \alpha} u^{\mu}  v^{\alpha} \\ 
&=& -\epsilon_{\alpha\mu} u^{\mu}  v^{\alpha} =
- u_{\alpha}  v^{\alpha}.
\end{eqnarray*}
Note that this shows 
we have a `see-saw rule' as long as a minus sign is introduced whenever a see-saw is performed.
The minus sign comes from the fact that the metric spinor $\epsilon_{\mu\nu}$ is antisymmetric
(it is a Levi-Civita symbol).
Setting $v^{\mu} = u^{\mu}$ we find
\[
u^{\mu} u_{\mu} = - u^{\mu} u_{\mu}
\]
and therefore $u^{\mu} u_{\mu} = 0$ as before.

We should expect the scalar invariant $u^{\mu} v_{\mu}$ to be something to do with the scalar
product of the associated 4-vectors, and you can confirm using  (\ref{get_flagpole}) that
\be
| u^{\mu} v_{\mu}|^2 = -\frac{1}{2} \ffU \cdot \ffV.        \gap{1} [ = |\biu^T \epsilon \biv|^2
\ee

A Hermitian matrix formed from a 4-vector as in
(\ref{XfromIxyz}) is of type $X^{\mu\bar{\nu}}$. Therefore the trace is not a Lorentz scalar
(that is, we can't set the two indices equal and sum).
To obtain a Lorentz scalar we can use $X^{\alpha \bar{\beta}} X_{\alpha \bar{\beta}}$.
More generally, for a pair of such spinors, you are invited to verify that
\be
X^{\alpha \bar{\beta}} Y_{\alpha \bar{\beta}}  = -2 \ffX \cdot \ffY.        \label{XdotYspinor}
\ee
This result makes it easy to convert some familiar tensor results into spinor notation.
For example, the continuity equation is
\be
\pp^{\alpha\bar{\beta}} J_{\alpha \bar{\beta}} = 0
\ee
where
\be
\pp^{\mu\bar{\nu}} = \sum_{\lambda} \pp^{\lambda} \sigma^{\lambda}
= \twomat -\frac{\partial}{c\partial t} + \frac{\partial}{\partial z} \; , &
 \frac{\partial}{\partial x} - i \frac{\partial}{\partial y} \\
 \frac{\partial}{\partial x} + i \frac{\partial}{\partial y}  \; , &
-  \frac{\partial}{c\partial t} - \frac{\partial}{\partial z}
\etwomat
 \ee
 and
 \be
 J^{\mu\bar{\nu}} = \twomat \rho c + j_z & j_x - i j_y \\ j_x + i j_y & \rho c - j_z \etwomat.
 \ee

Using (\ref{XdotYspinor}) again, the D'Alembertian can be written
\[
\Box^2 =  -\frac{1}{2} \pp^{\alpha \bar{\beta}} \pp_{\alpha \bar{\beta}}.
\]
You wouldn't ever want to write it like that, of course, since it is a scalar so you
may as well just write $\Box^2$ and convert it to $-(\partial/c\partial t)^2
+  (\partial/\partial x)^2
+  (\partial/\partial y)^2
+  (\partial/\partial z)^2$ when needed.

\else\omission\fi

\section{Applications}  \label{s.spinorapp}

\ifodd\final

We illustrate the application of spinors to something other than spin
by writing down Maxwell's equations in
spinor notation. This is merely to demonstrate that it can be done. We won't pursue whether
or not much can be learned from this, it is just to demonstrate that spinors are a flexible
tool.

To this end, introduce the quantity ${\bf F} = {\bf E} - i c{\bf B}$ where $\bf E$ and $\bf B$ are
the electric and magnetic fields. Form the mixed 2nd rank spinor
\be
F_{\bar{\mu}}^{\;\,\bar{\nu}} = \twomat F_z & F_x - i F_y \\ F_x + i F_y & - F_z \etwomat,
\ee
then Maxwell's equations can be written

\eqbox
{ }
{
\pp^{\mu \bar{\alpha}} F_{\bar{\alpha}}^{\;\,\bar{\nu}} =  c \mu_0 J^{\mu\bar{\nu}}.
}{spinormaxwell}

For example, the $\mu=1,\nu=1$ term on the left hand side is
\begin{eqnarray*}
-\frac{\partial F_z}{c\partial t} + \frac{\partial F_z}{\partial z}
+\frac{\partial F_x}{\partial x} + \frac{\partial F_y}{\partial y}
+i\left(\frac{\partial F_y}{\partial x} - \frac{\partial F_x}{\partial y}\right) \\
=  \Div {\bf F} - \left( \frac{\partial {\bf F}}{c\partial t} - i \Curl {\bf F} \right)_z .
\end{eqnarray*}
The real and imaginary parts of this are
\[
 \Div {\bf E} + c (\Curl {\bf B})_z - \frac{\partial E_z}{\partial t}  \;\;\; \mbox{and}\;\;
-c\Div {\bf B}  + (\Curl {\bf E})_z + \frac{\partial B_z}{\partial t}.
\]
Eq (\ref{spinormaxwell}) says the first of these is equal to $c \mu_0(\rho c+j_z)$, and
the second is equal to zero (since the $1,1$ term of the right hand side is real).
By evaluating the $2,2$ term you can find similarly that
\be
 \Div {\bf E} - c (\Curl {\bf B})_z + \frac{\partial E_z}{\partial t} &=& c \mu_0(\rho c - j_z)  \nonumber  \\
 \mbox{and}\;\; -c\Div {\bf B}  - (\Curl {\bf E})_z - \frac{\partial B_z}{\partial t} &=& 0 .
\ee
By taking sums of these simultaneous equations we find $\Div {\bf B}=0$ and
$\Div {\bf E} = \rho/\epsilon_0.$ By taking differences we find the $z$-component
of the other two Maxwell equations. You can check that the $1,2$ and $2,1$ terms of the spinor
equation yield the $x$ and $y$ components of the remaining Maxwell equations.

We now have a spinor-based method to 
obtain the transformation law for electric and magnetic fields: just transform
$F_{\bar{\mu}}^{\;\,\bar{\nu}}$. The result is exactly the same as one may obtain by
using tensor analysis to transform the Faraday tensor. It follows
that any antisymmetric second rank tensor can similarly be
`packaged into' a 2nd rank spinor whose indices are both of the same type.

A spinor for the 4-vector potential can also be introduced, and it is easy to write the
Lorenz gauge condition and wave equations, etc. The Lorentz force equation
is slightly more awkward---see exercises. Some solutions of Maxwell's equations can be
found relatively easily using spinors. An example is the radiation field due to an
accelerating charge: something that requires a long calculation using tensor methods.

\else\omission\fi

\section{Dirac spinor and particle physics}  \label{s.dirac_spinor}

We already mentioned in section (\ref{s.4vecspin}) that a pair of spinors can be
used to represent a pair of mutually orthogonal 4-vectors. A good way to do this
is to use a pair of spinors of opposite chirality, because then it is possible
to construct equations possessing invariance under parity inversion.
Such a pair is called
a {\em bispinor} or {\em Dirac spinor}. It can conveniently be written as a 4-component
complex vector, in the form
\be
\Psi = \twovec{\phi_R}{\chi_L}      \label{diracspinor}
\ee
where it is understood that each entry is a 2-component spinor, $\phi_R$ being right-handed
and $\chi_L$ left-handed. (Following standard practice in particle physics, we won't
adopt index notation for the spinors here, so the subscript $L$ and $R$ is introduced to keep track of
the chirality). Under change of reference frame $\Psi$ transforms as
\be
\Psi \rightarrow \twomat \Lambda(v) & 0 \\ 0 & \Lambda(-v) \etwomat \Psi  \label{transformDirac}
\ee
where each entry is understood to represent a $2 \times 2$ matrix, and we wrote
$\Lambda(v)$ for $\exp\left(i \boldsig \cdot \boldtheta/2 - \boldsig \cdot {\boldrho}/2 \right)$
and $\Lambda(-v)$ for $(\Lambda(v)^{\dagger})^{-1} =
\exp\left(i \boldsig \cdot \boldtheta/2 + \boldsig \cdot {\boldrho}/2 \right)$.
It is easy to see that the combination
\be
 \left( \phi_R^{\dagger}, \, \chi_L^{\dagger} \right)
 \twomat 0 & I \\ I & 0 \etwomat
\twovec{\phi_R}{\chi_L} =  \phi_R^{\dagger}\chi_L + \chi_L^{\dagger}\phi_R
          \label{dirac_invariant}
\ee
is Lorentz-invariant.

We will show how $\Psi$ can be used to represent the 4-momentum and 4-spin (Pauli-Lubanski 4-vector)
of a particle. First extract the 4-vectors given by the flagpoles of $\phi_R$ and $\chi_L$:
\be
\ffA^{\mu} = \bra{\phi_R} \sigma^{\mu} \ket{\phi_R}, \gap{0.6}
\ffB_{\mu} =  \bra{\chi_L} \sigma^{\mu} \ket{\chi_L}.  \label{ABmu}
\ee
Note that $\ffB_{\mu}$ has a lower index. This is because, in view of the fact that
$\chi_L$ is left-handed (i.e. negative chirality), under a Lorentz transformation
its flagpole behaves as a
covariant 4-vector. We would like to form the difference of these 4-vectors,
so we need to convert the second to contravariant form. This is done via the metric tensor
$g_{\mu\nu}$:
\[
\ffU^{\mu} = \left( \ffA^{\mu} - \ffB^{\mu} \right) =
\bra{\phi_R} \sigma^{\mu} \ket{\phi_R} - g^{\mu\alpha} \bra{\chi_L} \sigma^{\alpha} \ket{\chi_L}.
\]
(The notation is consistent if you keep in mind that the $_L$'s have lowered the index
on $\sigma$ in the second term).
In terms of the Dirac spinor $\Psi$, this result can be written as
\be
\ffU^0 = \Psi^{\dagger} \twomat I & 0 \\ 0 & I \etwomat \Psi
      = \phi_R^{\dagger} \phi_R  + \chi_L^{\dagger} \chi_L  \label{Uzero}
\ee
for the time component, and
\be
\ffU^i = \Psi^{\dagger} \twomat \boldsig & 0 \\ 0 & -\boldsig \etwomat \Psi   \label{Uspace}
\ee
for the spatial components.

Now introduce the $4 \times 4$ matrices, called {\bf Dirac matrices}:
\be
\gamma^0 = \twomat 0 & I \\ I & 0 \etwomat, \;\;
\gamma^i =  \twomat 0 & -\sigma^i \\ \sigma^i & 0 \etwomat.     \label{define_gammai}
\ee
Here we are writing these matrices in the `chiral' basis implied by the form
(\ref{diracspinor}); see exercise \ref{ex.Diracmatrices} for another representation.
Using these, we can write (\ref{Uzero}) and (\ref{Uspace}) both together, as
\be
\ffU^{\mu} = \Psi^{\dagger}  \gamma^0 \gamma^{\mu} \Psi.  \label{Ucomplete}
\ee
We now have a 4-vector extracted from our Dirac spinor. It can be of any type---it
need not be null. If in some particular reference frame it happens that
$\phi_R = \pm \chi_L$ (this equation is not Lorentz covariant so cannot be true in
all reference frames, but it can be true in one), then from eqn (\ref{ABmu})
we learn that in this reference frame the
components of $\ffA^{\mu}$ are equal to the components of
$\ffB_{\mu}$ so the spatial part of $\ffU$ is zero, while the time
part is not. Such a
4-vector is proportional to a particle's 4-velocity in its rest frame.
In other words, the Dirac spinor can be used to describe the 4-velocity of a massive
particle, and in this application it must have either
$\phi_R = \chi_L$ or $\phi_R = -\chi_L$ in the rest frame. In other frames the
4-velocity can be extracted using \eq{Ucomplete}.

Next consider the sum of the two flagpole 4-vectors. Let
\be
\ffW = m c S \left( \ffA + \ffB \right)
\ee
where $S$ is the size of the intrinsic angular momentum of the particle,
and introduce
\be
\gamma^5 = \twomat I & 0 \\ 0 & -I \etwomat.
\ee
By using
\be
\Sigma^{\mu} &=& \gamma^0 \gamma^{\mu} \gamma^5
 = \left( \twomat I & 0 \\ 0 & -I \etwomat, \; \twomat \boldsig & 0 \\ 0 & \boldsig \etwomat \right),
       \label{define_Sigmamu}
\ee
we can write
\be
\ffW^{\mu} = {m c S} \Psi^{\dagger} \Sigma^{\mu} \Psi.              \label{Wcomplete}
\ee
This 4-vector is orthogonal to $\ffU$. It can therefore be the 4-spin, if we choose
$\phi_R$ and $\chi_L$ appropriately. What is needed is that the spinor $\phi_R$
be aligned with the direction of the spin angular momentum in the rest frame. We already
imposed the condition that either $\phi_R = \chi_L$
or $\phi_R = - \chi_L$ in the rest frame, so it follows that either
both spinors are aligned with the spin angular momentum in the rest frame,
or one is aligned and the other opposed.

We now have a complete representation of the 4-velocity and 4-spin of a particle,
using a single Dirac spinor. A spinor equal to $(1,0,1,0)/\sqrt{2}$ in the rest frame,
for example, represents a particle with spin directed along the $z$ direction.
A spinor $(0,1,0,1)/\sqrt{2}$ represents a particle with spin in the $-z$
direction. More generally, $(\phi, \phi)/\sqrt{2}$ is a particle at rest with
spin vector $\phi^{\dagger} \boldsig \phi$. Table \ref{spin.t.spinstate}
gives some examples.

The states having $\phi_R = \chi_L$ in the rest frame cover half the available
state space; the other half is covered by $\phi_R = -\chi_L$ in the rest frame.
The spinor formalism is here
again implying that there may exist in Nature two types of particle,
similar in some respects (such as having the same mass), but not the same. 
In quantum field theory, it emerges that if the first set of states are particle
states, then the other set can be interpreted as antiparticle states. This
interpretation only emerges fully once we look at physical {\em processes},
not just states, and for that we need equations describing interactions and
the evolution of the states as a function of time---the equations of
quantum electrodynamics, for example. However, we can already note that
the spinor formalism is offering a natural language to describe a universe which can
contain both matter and anti-matter. The structure of the mathematics matches
the structure of the physics in a remarkable, elegant way. This inspires wonder
and a profound sense that there is more to the universe than a merely adequate
collection of ad-hoc rules.

As long as the spin and velocity are not exactly orthogonal,
in the high-velocity limit it is found that one of the two spinor components
dominates. For example, if we start from $\phi_R = \chi_L = (1,0)$ in the rest
frame, then transform to a reference frame moving in the positive $z$
direction, then
$\phi_R$ will shrink and $\chi_L$ will grow until in the limit $v \rightarrow c$,
$\phi_R \rightarrow 0$. This implies that a massless particle can be
described by a single (two-component) spinor, and we recover
the description we saw in section \ref{s.chirality} in connection with
the Weyl equations.  In particular, we find that a Weyl spinor has 
helicity of the same sign as its chirality.
(The example we just considered had negative
helicity because the spin is along $z$ but the particle's velocity is
in the negative $z$ direction in the new frame.)

\begin{table}
\begin{tabular}{cccl}
$\Psi$   & $\ffU$    & $2\ffW /(mcS)$ &  \\
$(1,0,1,0)/\sqrt{2}$ & $(1,0,0,0)$ & $(0,0,0,1)$ & at rest, spin up  \\
$(0,1,0,1)/\sqrt{2}$ & $(1,0,0,0)$ & $(0,0,0,-1)$ & at rest, spin down \\
$(1,1,1,1)/2$ & $(1,0,0,0)$ & $(0,1,0,0)$ & at rest, spin along $+x$  \\
$(1,-1,1,-1)/2$ & $(1,0,0,0)$ & $(0,-1,0,0)$ & at rest, spin along $-x$  \\
\hline
$(1,0,0,0)$   & $(1,0,0,1)$ & $(1,0,0,1)$ & $v_z=c$, +ve helicity \\
$(0,1,0,0)$   & $(1,0,0,-1)$ & $(1,0,0,-1)$ & $v_z=-c$, +ve helicity \\
$(0,0,1,0)$   & $(1,0,0,-1)$ & $(-1,0,0,1)$ & $v_z=-c$, $-$ve helicity \\
$(0,0,0,1)$   & $(1,0,0,1)$ & $(-1,0,0,-1)$ & $v_z=c$, $-$ve helicity
\end{tabular}
\caption{Some example Dirac spinors and their associated 4-vectors.}
\label{spin.t.spinstate}
\end{table}

A parity inversion ought to change the direction in space of $\ffU$ (since its spatial
part is a polar vector) but leave the
direction in space of $\ffW$ unaffected (since its spatial part is an axial vector).
You can verify that this is satisfied if the parity inversion is represented by the matrix
\be
P = \twomat 0 & I \\ I & 0 \etwomat.
\ee
The effect of $P$ acting on $\Psi$ is to swap the two parts, $\phi_R \leftrightarrow \chi_L$.
You can now verify that the Lorentz invariant given in (\ref{dirac_invariant}) is
also invariant under parity so it is a true scalar. The quantity
$\Psi^{\dagger} \gamma^0 \gamma^5 \Psi = \phi_R^{\dagger}\chi_L - \chi_L^{\dagger}\phi_R$
is invariant under Lorentz transformations but changes sign under parity, so is a pseudoscalar.
The results are summarised in table \ref{spin.t.vectors}.

\begin{table}
\begin{tabular}{ll}
$\Psi^{\dagger} \gamma^0 \Psi$  & scalar \\
$\Psi^{\dagger} \gamma^0 \gamma^5 \Psi$ & pseudoscalar \\
$\Psi^{\dagger} \gamma^0 \gamma^{\mu} \Psi$ & 4-vector $\ffU$, difference of flagpoles \\
$\Psi^{\dagger} \gamma^0 \gamma^{\mu} \gamma^5 \Psi$ & axial 4-vector $\ffW$, sum of flagpoles \\
$\Psi^{\dagger} \gamma^0 (\gamma^{\mu} \gamma^{\nu} - \gamma^{\nu} \gamma^{\mu}) \Psi$ & antisymmetric tensor
\end{tabular}
\caption{Various tensor quantities associated with a Dirac spinor. The notation
$\Psibar = \Psi^{\dagger} \gamma^0$ (called {\em Dirac adjoint})
can also be introduced, which allows
the expressions to be written $\Psibar \Psi$, $\Psibar \gamma^5 \Psi$, and so on.}
\label{spin.t.vectors}
\end{table}

\subsection{Moving particles and classical Dirac equation}

So far we have established that a Dirac spinor representing a massive particle
possessing intrinsic angular momentum ought to have $\phi_R = \chi_L$ in the rest
frame, with both spinors aligned with the intrinsic angular momentum. We have further
established that
under Lorentz transformations the Dirac spinor will continue to yield the
correct 4-velocity and 4-spin using eqs (\ref{Ucomplete}) and (\ref{Wcomplete}).

Now we shall
investigate the general form of a Dirac spinor describing a moving particle. All we
need to do is apply a Lorentz boost.
We assume the Dirac spinor $\Psi = (\phi_R(v), \chi_L(v))$
takes the form $\phi_R(0) = \chi_L(0)$
in the rest frame, and that it transforms as (\ref{transformDirac}).
Using (\ref{boostspin}) the Lorentz boost for a Dirac spinor can be written
\[
\Lambda = \cosh\! \left( \frac{\rho}{2} \right) \left( \begin{array}{cc}
 I - {\bf n} \cdot \boldsig \tanh(\rho/2) & 0 \\
0 & \rule{-3ex}{0pt} I + {\bf n} \cdot \boldsig \tanh(\rho/2) 
 \end{array} \right).
\]
Now (in units where $c=1$) $\cosh \rho = \gamma = E/m$ where $E$ is the energy of the particle,
and $\cosh \rho = 2 \cosh^2(\rho/2) - 1 = 2 \sinh^2(\rho/2) + 1$ so
\begin{eqnarray}
\cosh(\rho/2) &=& \left( \frac{E+m}{2m} \right)^{1/2} , \gap{1} \nonumber \\
\sinh(\rho/2) &=& \left( \frac{E-m}{2m} \right)^{1/2} ,    \label{coshsinh} \\
\tanh(\rho/2) &=& \left( \frac{E-m}{E+m} \right)^{1/2} = \frac{p}{E+m}. \nonumber
\end{eqnarray}
Therefore we can express the Lorentz boost in terms of energy
and momentum (of a particle boosted from its rest frame):
\be
\Lambda = \sqrt{ \frac{E+m}{2m} } \twomat I + \frac{\boldsig \cdot {\bf p}}{E+m} & 0 \\
0 & I - \frac{\boldsig \cdot {\bf p}}{E+m}   \etwomat
                \label{boostDirac}
\ee
where the sign is set such that ${\bf p}$ is the momentum of the particle in the new frame.

For example, consider the spinor $\Psi_0 = (1,0,1,0)/\sqrt{2}$, i.e. spin up along $z$ in the rest
frame. Then in any other frame,
\be
\Psi = \frac{1}{\sqrt{ 4m(E+m) }}\fourvec{E+m + p_z}{p_x + i p_y}{E+m - p_z}{-p_x-ip_y}
    \label{Psimoving}
\ee
(with $c=1$).
Suppose the boost is along the $x$ direction. Then, as $p_x$ grows larger,
$p_x \rightarrow E$, so the
positive chirality part has a spinor more and more aligned with $+x$, and the negative
chirality part has a spinor more and more aligned with $-x$. For a boost along $z$,
one of the chirality components vanishes in the limit $|v| \rightarrow c$.
This is the behaviour we previously discussed in relation to
table \ref{spin.t.spinstate}.

Next we shall present the result of a Lorentz boost another way. We will construct
a matrix equation satisfied by $\Psi$ that has the same form as the Dirac equation
of particle physics.
Historically, Dirac obtained his equation via a quantum mechanical argument.
However, the classical version can prepare us for the quantum
version, and help in the interpretation of the solutions.

We already noted that the Lorentz boost takes the form
\begin{eqnarray*}
\phi_R(\biv) &=& \left( I \cosh(\rho/2) - \boldsig \cdot {\bf n} \sinh(\rho/2) \right) \phi_R(0), \\
\chi_L(\biv) &=& \left( I \cosh(\rho/2) + \boldsig \cdot {\bf n} \sinh(\rho/2) \right) \chi_L(0).
\end{eqnarray*}
Using eq. (\ref{coshsinh}), and multiplying top and bottom by $(E+m)^{1/2}$, we find
\begin{eqnarray*}
\phi_R({\bf p}) &=& \frac{E+m + \boldsig \cdot {\bf p}}{[2m(E+m)]^{1/2} } \phi_R(0), \gap{1}\\
\chi_L({\bf p}) &=& \frac{E+m - \boldsig \cdot {\bf p}}{[2m(E+m)]^{1/2} } \chi_L(0).
\end{eqnarray*}
where ${\bf p} = - \gamma m\vv$ is the momentum of the particle in the new frame.
(The same result also follows immediately from eqn (\ref{boostDirac}).)
Introducing the assumption $\phi_R(0) = \chi_L(0)$ we obtain from these two equations,
after some algebra\footnote{Let $\eta=\boldsig \cdot {\bf p}$.
Premultiply the first equation by $E+m-\eta$
and the second by $E+m+\eta$ to obtain
$(E+m-\eta)\phi_R = (E+m+\eta)\chi_L$; then premultiply the first equation by
$E-m-\eta$ and the second by $-E+m-\eta$ to obtain
$(E-m-\eta)\phi_R = (-E+m-\eta)\chi_L$ (after making use of $\eta^2 = p^2$). The sum and
difference of these equations gives the result.},
\ben
( E - \boldsig \cdot {\bf p} )\phi_R({\bf p}) &=& m \chi_L({\bf p}), \\
( E + \boldsig \cdot {\bf p} )\chi_L({\bf p}) &=& m \phi_R({\bf p}).
\een
The left hand sides of these equations are the same as in the Weyl equations;
the right hand sides have the requisite chirality. As a set, this
coupled pair of equations is parity-invariant, since under a parity inversion
the sign of $\boldsig \cdot {\bf p}$ changes and $\chi$ and $\phi$ swap over.
In matrix form the equations can be written
\be
\twomat   E- \boldsig \cdot {\bf p}  & -m \\
-m &  E + \boldsig \cdot {\bf p}   \etwomat \twovec{\phi_R({\bf p})}{\chi_L({\bf p})} = 0.
    \label{classicalDirac}
\ee
This equation is very closely related to the Dirac equation.
One may even go so far as to say that
(\ref{classicalDirac}) ``is'' the Dirac equation in free space, if one re-interprets
the terms, letting $E$ and $p$ be the frequency and wave-vector of a plane wave, up
to factors of $\hbar$.
In the present context
the equation represents a constraint that must be satisfied by any Dirac spinor
that represents 4-momentum and intrinsic spin of a given particle.

If we premultiply (\ref{classicalDirac}) by $\gamma^0$ then we have the form
\be
\twomat -m & E + \boldsig \cdot {\bf p} \\
E - \boldsig \cdot {\bf p} & -m  \etwomat \twovec{\phi_R({\bf p})}{\chi_L({\bf p})} = 0.
    \label{classicalDirac2}
\ee
This can conveniently be written
\be
( - \gamma^\lambda P_\lambda - m ) \Psi = 0
\ee
(see also exercise \ref{ex.Hamiltonian}.)

Our discussion has been entirely classical (in the sense of not quantum-mechanical).
In quantum field theory the spinor plays a central role. One has a spinor field, the excitations
of which are what we call spin $1/2$ particles. The results of this section reemerge
in the quantum context, unchanged for energy and momentum eigenstates, and
in the form of mean values or `expectation values' for other states.



\subsection{The standard representation}

In order to present Dirac spinors, we found it helpful to write them in component form as
in eqn (\ref{diracspinor}). This amounts to choosing a basis. The choice we made is
called the {\em chiral basis}. It is the most natural choice in which to discuss chirality,
and it gives the convenient fact that in this basis the Lorentz transformation matrix
((\ref{transformDirac}) and (\ref{boostDirac})) is block-diagonal. In the application
to particle physics, especially in the case of slow-moving particles, another basis is convenient.
This is called the `standard representation' or `Dirac representation', whose basis
vectors are related to the chiral basis by the transformation
\[
U = \frac{1}{\sqrt{2}} \twomat I & I \\ I & -I \etwomat.
\]
For example, in the standard representation, $\Psi$ as given in eqn (\ref{diracspinor}) would be 
written
\[
\Psi = \frac{1}{\sqrt{2}} \twovec{\phi_R + \chi_L}{\phi_R - \chi_L}.
\]
For a particle with spin described by a spinor $\psi$ in the rest frame, we have
$\phi_R = \chi_L = \psi / \sqrt{2}$ in the rest frame, and therefore in other frames,
\begin{eqnarray*}
\phi_R &=& (E+m+  \boldsig \cdot \vp) \psi (4m(E+m))^{-1/2}, \\
\chi_L &=& (E+m -  \boldsig \cdot \vp) \psi (4m(E+m))^{-1/2}, 
\end{eqnarray*}
using
eqn (\ref{boostDirac}). Thus, when expressed in the standard representation, the resulting Dirac spinor
is
\be
\Psi = \frac{1}{\sqrt{2m(E+m)}} \twovec{ (E+m) \psi }{\vp \cdot \boldsig \psi}.
\ee
For low velocities, $v \ll c$, we have $E = m + O(v^2)$ and hence, to first
order in $v$,
\be
\Psi  \simeq  \twovec{ \psi }{ \vv \cdot \boldsig \psi /2}.       \label{Psilowv}
\ee

\subsection{Electromagnetic interactions and $g=2$}

Introduce the matrices $\beta \equiv \gamma^0$ and $\alpha^i \equiv \gamma^0 \gamma^i$.
A useful way to write the Dirac equation (\ref{classicalDirac2}) is (exercise \ref{ex.Hamiltonian})
\be
H \Psi = E \Psi          \label{HPsi}
\ee
where $H = \boldalpha \cdot \vp + \beta m$. 
Eqn (\ref{HPsi}) suggests
that we should regard $H$ as a Hamiltonian. The standard way to treat the motion of a charged
particle in an electromagnetic field
in special relativity is to add a potential energy term $q\phi$ to the Hamiltonian, and replace
$\vp$ in the Hamiltonian by  $\tilde{\vp} - q\vA$ where $\vA$ is the vector potential and $\tilde{\vp}$
is the canonical momentum \cite{12Steane}. It is standard practice, in quantum mechanics and particle physics,
to use the symbol $\vp$ for canonical momentum, so we shall adopt that notation, and use
the symbol $\vp_{\rm k}$ for the kinetic momentum, such that
\[
\vp_{\rm k} = \vp - q \vA = \gamma m \vv .
\]
We thus obtain the Hamiltonian
\be
H = \boldalpha \cdot (\vp - q \vA) + \beta m + q \phi .
\ee
In the standard representation, eqn (\ref{HPsi}) now reads
\be
\twomat E-V-m &  - \boldsig\cdot(\vp-q\vA) \\
-\boldsig\cdot(\vp-q\vA) & E-V + m \etwomat   \twovec{\psi_+}{\psi_-} = 0 \label{Diracem}
\ee
where we introduced $\psi_{\pm} \equiv (\phi_R \pm \chi_L)/\sqrt{2}$ and the potential
energy $V=q\phi$. 

Let us treat motion in a pure magnetic field, so $V=0$. The second row of (\ref{Diracem})
tells us that 
\[
\psi_- = \frac{\boldsig \cdot \vp_{\rm k}}{E+m} \psi_+ \simeq  \frac{\boldsig \cdot \vp_{\rm k}}{2m} \psi_+
\]
where the second version is valid at low speeds (c.f. eqn (\ref{Psilowv})). Since $p \simeq mv$
at $v \ll 1$,  we then have that $|\phi_-| \simeq (v/2) |\psi_+|$. For this reason $\psi_+$ and $\psi_-$ are
called the `large' and `small' components in this context. The first row of (\ref{Diracem}) gives
\be
(E-m)\psi_+ = \boldsig\cdot \vp_{\rm k} \psi_- 
\simeq \frac{(\boldsig\cdot \vp_{\rm k}) (\boldsig\cdot \vp_{\rm k})}{2m} \psi_+  .
\label{psipluseq}
\ee
Now using the identity
\be
(\boldsig \cdot \va)(\boldsig \cdot \vb) = \va \cdot \vb + i \boldsig\cdot \va\wedge\vb
\ee
we have 
\be
(\boldsig\cdot \vp_{\rm k}) (\boldsig\cdot \vp_{\rm k}) = p_{\rm k}^2 + i \boldsig \cdot
(\vp - q\vA) \wedge (\vp - q\vA). \;\;
\ee
In classical physics, the second term here would be zero, but in quantum physics it is not.
Although our whole presentation has been classical up till now, we will now make a small
foray into quantum mechanics. We regard $\vp$ has an operator, which in the
position representation is expressed $\vp = -i\hbar \Grad$, and now the spinor $\psi_+$ has
to be thought of as a function of position---it is a wavefunction with two components.
We have
\be
[ (\vp - q\vA) \! &\wedge & \!(\vp - q\vA) ] \psi_+  \nonumber\\
&=& -i\hbar q\left( \Curl (\vA\psi_+) + \vA \wedge (\Grad\psi_+) \right)  \nonumber\\
&=& -i \hbar q (\Curl \vA) \psi_+
\ee
so eqn (\ref{psipluseq}) reads
\be
(E-m) \psi_+ = \left( \frac{(\vp - q\vA)^2}{2m} + \frac{\hbar q}{2 m} \boldsig \cdot \vB \right) \psi_+.
\ee
This is the time-independent Schr\"odinger equation for a particle interacting with a magnetic field,
such that the operator $(-\hbar q /2m)\boldsig$ represents the magnetic dipole moment of
the particle. One finds that the angular momentum of the particle is represented by the operator
$\boldsig \hbar/2$ (exercise \ref{ex.spinDirac}), so the 
gyromagnetic ratio is $g=2$. Thus the value of the $g$-factor of a spin half
particle described by the Dirac equation is not an independent variable but is constrained to
take the value 2.

\section{Spin matrix algebra (Lie algebra)*}  \label{s.spin_algebra}

We introduced the Pauli spin matrices abruptly at the start
of section \ref{s.su2}, by giving a set of matrices and their commutation relations.
By now the reader has some idea of their usefulness. 

In group theory, these matrices are called the {\em generators} of the group SU(2), because
any group member can be expressed in terms of them in the form $\exp(i\boldsig \cdot \boldtheta/2)$.
More precisely, the Pauli matrices are the generators of one {\em representation} of the
group SU(2), namely the representation in terms of $2 \times 2$ complex matrices. Other
representations are possible, such that there is an isomorphism between one representation
and another. Each representation will have generators in a form suitable for that
representation. They could be matrices of larger size, for example, or even differential
operators. In every representation, however, the generators will have the same behaviour
when combined with one another, and this behaviour reveals the nature of the group. This means
that the innocent-looking commutation relations
(\ref{sigmacommute}) contain much more information than one might have supposed:
they are a `key' that, through the use of $\exp(i\boldsig \cdot \boldtheta/2)$, unlocks
the complete mathematical behaviour of the group. In Lie group theory these equations
describing the generators are called the `Clifford algebra' or `Lie algebra' of the group.

If a Lie group does not have a matrix representation it can be hard or impossible to give
a meaningful definition to $\exp(M)$ where $M$ is a member of the group. In this case
one uses the form $I + \epsilon M$ to write a group member infinitesimally close
to the identity for $\epsilon \rightarrow 0$. The generators are a subgroup
such that any member close to $I$ can be written $I + \epsilon G$ where $G$ is in the
generator group. This is the more general definition of what is meant by
the generators.

The generators of rotations in three dimensions (\ref{Jgenerator}) have the commutation relations
\be
\left[J_x,\, J_y\right] = i J_z  \;\; \mbox{ and cyclic permutations.}
\ee
By comparing with (\ref{sigmacommute}) one can immediately deduce the relationship
between SU(2) and SO(3), including the angle doubling!

The restricted Lorentz group has generators $K_i$ (for boosts) and $J_i$ (for rotations).
A matrix representation (suitable for a rectangular coordinate system) is
\be
J_x = \fourmat
0&0&0&0\\ 0&0&0&0 \\ 0&0&0&-i \\ 0&0&i&0 \efourmat, \;\;\;\;\;
K_x = \fourmat
0&i&0&0\\ i&0&0&0 \\ 0&0&0&0 \\ 0&0&0&0 \efourmat,   \label{Kxmatrix}  \\
J_y = \fourmat
0&0&0&0\\ 0&0&0&i \\ 0&0&0&0 \\ 0&-i&0&0 \efourmat, \;\;\;\;\;
K_y = \fourmat
0&0&i&0\\ 0&0&0&0 \\ i&0&0&0 \\ 0&0&0&0 \efourmat, \\
J_z = \fourmat
0&0&0&0\\ 0&0&-i&0 \\ 0&i&0&0 \\ 0&0&0&0 \efourmat, \;\;\;\;\;
K_z = \fourmat
0&0&0&i\\ 0&0&0&0 \\ 0&0&0&0 \\ i&0&0&0 \efourmat.           \label{SKdef}
\ee
The commutation relations are (with cyclic permutations)
\begin{eqnarray*}
\left[J_x, \, J_y\right] &=& i J_z \\
\left[K_x, \, K_y\right] &=& - i J_z \\
\left[J_x, \, K_x\right] &=& 0 \\
\left[J_x, \, K_y\right] &=& i K_z \\
\left[J_x, \, K_z\right] &=& i K_y.
\end{eqnarray*}
The second result shows that the Lorentz boosts on their own do not form a closed
group, and that two boosts can produce a rotation: this is the Thomas precession.
If we now form the combinations
\[
{\bf A} = ({\bf J} + i {\bf K})/2, \gap{2}
{\bf B} = ({\bf J} - i {\bf K})/2, \gap{2}
\]
then the commutation relations become
\begin{eqnarray*}
\left[A_x, \, A_y\right] &=& i A_z \\
\left[B_x, \, B_y\right] &=& i B_z \\
\left[A_i, \, B_j\right] &=& 0, \;\;\; (i,j = x,y,z).
\end{eqnarray*}
This shows that the Lorentz group can be divided into two groups, both SU(2),
and the two groups commute. This is another way to deduce the existence
of two types of Weyl spinor and thus to define chirality.

The Clifford algebra satisfied by the
Dirac matrices $\gamma^0$, $\gamma^1$, $\gamma^2$, $\gamma^3$ is
\be
\left\{ \gamma^{\mu}, \gamma^{\nu} \right\} = -2 \eta^{\mu\nu} I       \label{gamma_anticomm}
\ee
where $\eta^{\mu\nu}$ is the Minkowski metric and $I$ is the unit matrix. That is,
$\gamma^0$ squares to $I$ and $\gamma^i$ ($i=1,2,3$) each square to $-I$, and they all anticommute
among themselves. These anti-commutation relations are normally taken to be the {\em defining property}
of the Dirac matrices. A set of quantities $\gamma^{\mu}$ satisfying such anti-commutation relations
can be represented using $4 \times 4$ matrices in more than one way---c.f. exercise \ref{ex.Diracmatrices}. 
If the metric of signature $(1,-1,-1,-1)$ is used, then
the minus sign on the right hand side of (\ref{gamma_anticomm}) becomes a plus sign.

\subsection{Dirac spinors from group theory*}

\ifodd\final

We conclude with a demonstration of how to establish the main properties of Dirac spinors
by using group theory.

First let's take a fresh perspective on the six generators of the Lorentz group
in 4-dimensional spacetime, i.e. the $J_i$ and $K_i$ matrices defined in eqs.
(\ref{Kxmatrix})--(\ref{SKdef}). There are six entities, three of which are used
to form a polar vector, and three an axial vector. You guessed it: we can
gather them together into antisymmetric tensor $\LM^{\mu\nu}$. This is a tensor
of matrices, or, to be more general, of objects that behave like matrices
having given commutation relations. You can check that the $J_i$ and $K_i$ matrices can all be written
\[
(\LM^{ab})^{c}_{\;d} =
\eta^{bc} \delta^{a}_{d}
- \eta^{ac} \delta^{b}_{d}
\]
where by picking the 6 combinations
$(a,b) = (0,1),\;(0,2),\;(0,3),\;,(1,2),\;(1,3),\;(2,3)$ the expression gives the
6 matrices. For example, $\LM^{01}$ is $-iK_x$, $\LM^{12}$ is $-iJ_z$, etc.
We can now write any Lorenz transformation as
\[
\Lambda = \exp\left( \half \theta_{\mu\nu} \LM^{\mu\nu} \right)
\]
where, don't forget, each $\LM^{ab}$ is a $4\times 4$ matrix. $\theta_{ab}$
provides 6 numbers telling which transformation we want.

These generators of the Lorentz group obey the following Clifford algebra, which is called
the {\em Lorentz Lie algebra}:
\be
\left[\LM^{\mu\nu}, \LM^{\rho\sigma}\right]
=
 \eta^{\mu\rho} \LM^{\nu\sigma}
\! - \eta^{\nu\rho} \LM^{\mu\sigma}
\! + \eta^{\nu\sigma} \LM^{\mu\rho} 
\! - \eta^{\mu\sigma} \LM^{\nu\rho} . \nonumber \\
\ee

Now we can connect the Clifford algebra to the Lorentz group. Let
\[
\ftS^{\mu\nu} = \frac{1}{4} [ \gamma^{\mu}, \; \gamma^{\nu} ] = \frac{1}{4}
(\gamma^{\mu} \gamma^{\nu} - \gamma^{\nu}, \; \gamma^{\mu}),
\]
then {\em the matrices $S^{\mu\nu}$ form a representation of the Lorentz algebra}:
\[
\left[\ftS^{\mu\nu}, \ftS^{\rho\sigma}\right]
= \eta^{\nu\rho} \ftS^{\mu\sigma}
- \eta^{\mu\rho} \ftS^{\nu\sigma}
+ \eta^{\mu\sigma} \ftS^{\nu\rho}
- \eta^{\nu\sigma} \ftS^{\mu\rho} .
\]
{\em Proof}: exercise for the reader! You may find it helpful first to obtain
\ben
\ftS^{ab} &=& \half (\gamma^{a} \gamma^{b} + \eta^{ab}), \\
\left[\ftS^{ab}, \gamma^{c}\right] &=& \gamma^{b}\eta^{ca} - \gamma^{a} \eta^{bc} .
\een

It follows that if we introduce an object $\psi$ that can be acted upon by the matrices $\ftS$, then
we shall be able to Lorentz-transform it using
\[
\psi \rightarrow \exp\left( \half \theta_{\mu\nu} \ftS^{\mu\nu} \right) \psi.
\]
The object $\psi$ can be a set of four complex numbers. It is a Dirac spinor.

Let
\[
\tilde{\Lambda} \equiv \exp\left( \half \theta_{\mu\nu} \ftS^{\mu\nu} \right)
\]
(the tilde is to distinguish this from the Lorentz transformation of 4-vectors). You can
prove that
\[
\tilde{\Lambda}^{\dagger} = \gamma^0 \tilde{\Lambda}^{-1} \gamma^0.
\]
Then with some further effort one can obtain central properties such as the Lorentz invariance of
$\psi^{\dagger} \gamma^0 \psi$, and that $\psi^{\dagger} \gamma^0 \gamma^{\mu} \psi$
is a 4-vector, etc.

\else\omission\fi

\ifodd\final

\section{Exercises}



\begin{enumerate}

\item  \label{ex.expsig}
Show that the Pauli matrices all square to 1, i.e. $\sigma_x^2 = \sigma_y^2 = \sigma_z^2 = I$.
Hence, using $\exp(M) \equiv \sum_n M^n/n!$, prove eq. (\ref{expsigma}).

\item \label{ex.unitary}
(a) Prove that any SU(2) matrix can be written
$a I + i b \sigma_x + i c \sigma_y + i d \sigma_z$
where $a,b,c,d$ are real and $a^2 + b^2 + c^2 + d^2 = 1$. [e.g. start from an arbitrary
$2\times 2$ matrix $M$ and show that if $M^{-1}=M^\dagger$
and $|M|=1$ then $M_{11}=M_{22}^*$ and $M_{12} = -M_{21}^*$].\\
(b) Show that any SU(2) matrix can be
written $\exp(i \boldtheta \cdot \boldsig/2)$. [e.g. prove that this
form always gives an SU(2) matrix and spans the space].

\item Show that $g_{\mu\alpha}(\biu^{\dagger} \sigma^{\alpha} \biu)
= (\epsilon \biu^*)^{\dagger} \sigma^{\mu} (\epsilon \biu^*)$ and interpret this
result (c.f. eqs (\ref{get_flagpole}) and (\ref{Vfrom_tildeu}), and the caption
to table \ref{spin.t.trans}). [Method: either manipulate the matrices, or just
write $\biu = (a,b)$ and evaluate all the terms].

\item Find the flagpole 4-vectors for the following spinors, and confirm that they are null:
$(1,1), \; (-2,1), \; (2,1+i)$.\\
{\em Ans} (2,2,0,0); $(5,-4,0,3)$; $(6,4,4,2)$.

\item Prove the statement after eqn (\ref{zboostspin}).

\item Starting from eqs. (\ref{Ucomplete}) and (\ref{Wcomplete}), show that the effect
of swapping $\phi_R$ and $\chi_L$ is to change the sign of the spatial part of $\ffU$ and
the time part of $\ffW$.

\item
Starting from eqs. (\ref{Ucomplete}) and (\ref{Wcomplete}),
confirm that $\ffU$ and $\ffW$ are orthogonal.

\item Bearing in mind the commutation relations for Pauli matrices, show that, for
any 3-vector ${\bf w}$, $(\boldsig \cdot {\bf w})^2 = w^2$, and hence complete the
steps leading to the Dirac equation (\ref{classicalDirac}).

\item An electron moving along the $x$ axis with speed $0.8c$ has its spin in the
$(1,0,1)$ direction in the lab frame. Adopting units where $c=1$ and the size of the spin is
$1/2$, construct a Dirac spinor appropriate to describe
this electron in the lab frame. [Method: first obtain the 4-velocity $\ffU$
and 4-spin $\ffW$, hence obtain the two flagpoles
and hence the two spinors]. Transform this spinor to the rest frame, and find
the direction of the spin in the rest frame. 

\item
Find the two null 4-vectors (flagpoles) associated with the
Dirac spinor $\Psi = (1.17005,\;  0.204124,\;  0.462943,\;  -0.204124)$.
Assuming this spinor represents the motion of a particle, find
the 4-velocity and the direction of the spin in the rest frame.

\item
Investigate
$q F^{\mu}_{\;\,\alpha} U^{\alpha \bar{\nu}}$ where $F$ is the electromagnetic field spinor
and $U^{\alpha \bar{\nu}}$ is
the 4-velocity spinor. The result is not simply
the Lorentz force, but can be understood in terms of the Lorentz force and
a force obtained from the dual of the Faraday tensor:
\[
i c f^{\mu \bar{\nu}} + c \tilde{f}^{\mu \bar{\nu}} = -q F^{\mu}_{\;\,\alpha} U^{\alpha \bar{\nu}}
\]
[To obtain $F^{\mu}_{\;\,\nu}$ from $F_{\bar{\mu}}^{\;\,\bar{\nu}}$, first take the complex conjugate
then raise and lower indices. One finds that the matrix is unchanged except the sign of ${\bf E}$
is reversed.]

\item \label{ex.Hamiltonian}
Let $\beta \equiv \gamma^0$ and $\alpha^i \equiv \gamma^0 \gamma^i$.
Show that, using these matrices, the Dirac equation (\ref{classicalDirac2}) may be written $H \Psi = E \Psi$ where
$H = \boldalpha \cdot \vp + \beta m$.

\item \label{ex.Diracmatrices}
Show that, in the standard representation, the Dirac matrices take the form
\[
\gamma^0 = \twomat I & 0 \\ 0 & -I \etwomat, \;\;\;
\gamma^i = \twomat 0 & \sigma^i \\ -\sigma^i & 0 \etwomat .
\]

\item \label{ex.spinDirac}
In quantum mechanics, the Dirac equation forces us to conclude that the operator representing
spin angular momentum is $(\hbar/2)\boldsig$, not some other multiple. Prove this, as follows.
Treat the Dirac equation in free space, so $H =  \boldalpha \cdot \vp + \beta m$.
Let $\vJ \equiv \vL + \vS$ be the total angular momentum operator, with $\vL \equiv \vr \wedge \vp$
and $\vS$ to be discovered. Show that $[\vL, H] = i \hbar \boldalpha \wedge \vp$
(e.g. treat just the $z$ component and the others follow). Clearly, this is non-zero so we do
not have overall rotational invariance of the energy unless $\vS$ also contributes to the angular
momentum, such that $[\vS,H] = - i \hbar \boldalpha \wedge \vp$. Verify that the solution is
\[
\vS = \frac{\hbar}{2} \twomat \boldsig & 0 \\ 0 & \boldsig \etwomat .
\]

\end{enumerate}

\else\omission\fi

\bibliography{spinorrefs}

\end{document}